\documentclass[
	a4paper, 
	10pt, 
	unnumberedsections, 
	twoside, 
]{LTJournalArticle}

\PassOptionsToPackage{table}{xcolor}
\usepackage{xcolor}
\usepackage{subcaption}
\usepackage{tabularx}

\definecolor{revisionColor}{RGB}{255,0,0}
\definecolor{cleanColor}{RGB}{0,0,0}
\colorlet{changeColorVar}{cleanColor} 

\runninghead{Visual and Cognitive Biases in Scatterplot Interpretations} 

\footertext{} 

\makeatletter
\def\thanks#1{\protected@xdef\@thanks{\@thanks
        \protect\footnotetext{#1}}}
\makeatother

\title{Visual Elements and Cognitive Biases Influence Interpretations of Trends in Scatter Plots} 

\author{%
	Alexandre Filipowicz\ddag,
    Scott Carter\ddag,
    Nayeli Bravo,
    Rumen Iliev,
    Shabnam Hakimi,\\
    David A. Shamma,
    Kent Lyons,
    Candice Hogan \&
    Charlene Wu
    \thanks{\ddag Alexandre Filipowicz and Scott Carter contributed equally to this research. Corresponding author: Alexandre Filipowicz \href{mailto:alex.filipowicz@tri.global}{alex.filipowicz@tri.global}}
}

\date{\footnotesize Toyota Research Institute, Los Altos, CA, USA}

\renewcommand{\maketitlehookd}{%
	\begin{abstract}
		\noindent Visualizations are common methods to convey information but also increasingly used to spread misinformation. It is therefore important to understand the factors people use to interpret visualizations. In this paper, we focus on factors that influence interpretations of scatter plots, investigating the extent to which common visual aspects of scatter plots (outliers and trend lines) and cognitive biases (people’s beliefs) influence perceptions of correlation trends. We highlight three main findings: outliers skew trend perception but exert less influence than other points; trend lines make trends seem stronger but also mitigate the influence of some outliers; and people’s beliefs have a small influence on perceptions of weak, but not strong correlations. From these results we derive guidelines for adjusting visual elements to mitigate the influence of factors that distort interpretations of scatter plots. We explore how these guidelines may generalize to other visualization types and make recommendations for future studies.
	\end{abstract}
}

\begin{document}

\maketitle 


\section{Introduction}

Since William Playfair introduced the graphical display of statistical information~\cite{Playfair1801}, there has been a large increase in the use of graphs, charts, and other information visualization approaches. Statistical visualizations are now not only, a ``universal language''~\cite{Funkhouser1937}, but also a method of communication nearly as ubiquitous as language itself. Furthermore, although data visualizations are often created with the assumption that they represent unambiguous, unadulterated facts, they are in fact, much like language, subject to interpretation. 

This fact has become ever present in today's information age, where graphs are shared widely and easily across many platforms. Visualizations can provide a powerful way to summarize and convey complex information and are more effective at persuading viewers than are other representations of data (e.g., tables~\cite{pandey2014persuasive}). However, although the growing access to information has a number of advantages, it has also had the unintended consequence of facilitating the spread of misinformation~\cite{lazer2018science}. Additionally, graph literacy is not ubiquitous: even well educated people can have difficulty interpreting basic linear graphs~\cite{Bragdon2019UniversitySG}. Thus, the utility of graphical information depends largely on how they are perceived. 

The goal of the current article is to study the influence of different factors in people's interpretations of graphical trends in scatter plots. We focus specifically on scatter plots as these are among the most commonly used visualizations~\cite{battle2018beagle} and are particularly open to subjective interpretation because they represent actual data points (rather than data measures such as quantities or averages) that the observer assembles into meaningful trends and representations.

Accurate interpretations of trends in scatter plots can be strongly influenced by a plot's visual features~\cite{8305493}. Some of these features relate to low-level visual properties---such as the markers, colors, and textures~\cite{harrison2014corrweber, smart2019measuring, woodin2021conceptual, li2009evaluation, wang2022makes}---whereas others include higher-level features, such as the title or aspect ratio~\cite{WANZER2021101896, 2006-banking, Cairo2019}. It is therefore important to understand the extent to which these visual features influence people's interpretations of graphical information~\cite{9195155}.

In addition to visual features, people's beliefs and expectations can also affect how they interpret visual information. People are well known to exhibit ``confirmation biases'', tending to focus on information that confirms their existing beliefs while down-weighting information that goes against their beliefs~\cite{nickerson1998confirmation}. Indeed, our beliefs play a pivotal role in our perception of visual information. People tend to be less accurate at remembering graphical trends when these reflect gaps in people's knowledge~\cite{kim2017explaining}; having people explicitly identify these knowledge gaps can help them better integrate new information, even if it contradicts their prior knowledge~\cite{kim2017explaining}.  Research in cognitive psychology also consistently finds that prior beliefs can bias how people interpret low-level visual information (e.g., orientations~\cite{jabar2017tuned}, spatial
position~\cite{glaze2015normative}, movement~\cite{gold2007neural,glaze2015normative}, and object size~\cite{stottinger2006dissociating}). In the domain of visualizations, graphs play a central role in spreading misinformation, either by manipulating ways in which the visual elements of a graph are presented or playing to people's existing beliefs (e.g., to spread falsehoods about COVID-19~\cite{lee2021viral} or climate change~\cite{cook2020introduction}). Therefore, in addition to visual graph features, it is important to understand how beliefs and biases influence graphical interpretations, and how these potentially interact with visual features.

This paper contributes four experiments designed to explore how both common visual features and people's personal beliefs impact the interpretation of correlations presented in scatter plots. Experiments 1 and 2 examine the influence of outliers and trend lines on people's interpretations of correlation trends. Experiments 3 and 4 examine the influence of beliefs and their interaction with outliers on interpretations of correlations that either confirm or go against people's pre-existing beliefs. We use the results of these studies to suggest a set of guidelines that designers can use when developing visualizations.

\section{Related Work}
In general, people can accurately perceive scatter plot trends~\cite{correll2017regression, harrison2014corrweber, liu2021data}. For example, Correll and Heer found that people can draw accurate regression lines that match both simple linear and more complex non-linear bi-variate relationships~\cite{correll2017regression}. However, there are a number of visual elements that can influence this accuracy. For example, properties of the markers used to represent individual points can bias trend estimates, either by changing the salient aspects of the plot through differences in color~\cite{smart2019measuring}, size~\cite{woodin2021conceptual} or contrast~\cite{li2009evaluation, woodin2021conceptual}, or by biasing perceptions of trend directions through marker shape~\cite{liu2021data,tremmel1995visual,smart2019measuring}. To combat some of these perceptual biases, frameworks have been created to adjust the visual display of scatter plots using automated tools that account for visual and perceptual limitations~\cite{micallef2017towards}.

Properties of the data being plotted can also influence people's perception of trends. Outliers, or points falling far from a general trend, reliably bias people's interpretations in the direction of the outliers ~\cite{liu2021data,correll2017regression,ciccione2022outlier}. This influence can be quite robust, even when users are explicitly instructed to exclude outliers in their estimate of an overall trend~\cite{ciccione2022outlier}. However, although outliers do exert some influence on people's estimates, previous research also suggests that they are given less weight than points that are part of the main trend~\cite{correll2017regression, liu2021data, ciccione2022outlier}. These results have been attributed to the human propensity to code visual information into consistent \emph{sets} or \emph{ensembles} that exclude highly deviant points~\cite{szafir2016four, ariely2001seeing}. These findings are also consistent with principles of \emph{efficient coding} from psychology and neuroscience, such that perceptual processes weight consistent ensembles of information more heavily than outlying data points~\cite{summerfield2015humans, filipowicz2018rejecting, wei2015bayesian, de2011robust}.

However, there are still some open questions regarding the generalizability of these findings. The influence of outliers has generally been studied in contexts where participants use and adjust a trend line to indicate trend estimates~\cite{correll2017regression, liu2021data, ciccione2022outlier}. Although trend lines are a straightforward means to elicit participant responses, it is unclear to what extent this type of modality influences people's perceptions of scatter plot elements. Trend lines are a highly salient annotation method that can dominate people's attention when interpreting scatter plots~\cite{wang2022makes,8305493} and can influence trend perception independent of the underlying data points~\cite{reimann2021visual}. As such, Experiments 1 and 2 in the current article examine the generalizability of previous work with outliers, as well as the influence of trend lines on people's perceptions of trends in scatter plots.

In addition to perceptual factors, cognitive biases can also influence how people interpret information presented in scatter plots. Dimara and colleagues provide an extensive taxonomy of the different cognitive biases identified by decades of psychological research and categorize these biases based on the ``tasks'' these might influence~\cite{dimara2018task}. These tasks cover different times in which cognitive biases may exert an influence related to visualizations, such as when a person is~\emph{estimating} a trend in a graph, making a~\emph{decision}, or~\emph{recalling} information contained within a graph (see~\cite{dimara2018task} for seven such tasks). In addition to highlighting some of these biases, systems have also been proposed to both measure when cognitive biases may be influencing a person's perception of a scatter plot and propose interventions that specifically target the detected bias~\cite{wall2017warning}.

The current article investigates the influence of participant beliefs in the way scatter plot trends are perceived (what Dimara and colleagues would term an ``estimation'' task~\cite{dimara2018task}). Research in cognitive psychology consistently demonstrates that cognitive biases related to beliefs can influence human perception of aspects such as judgements of orientations~\cite{jabar2017tuned}, spatial position~\cite{glaze2015normative}, movement~\cite{gold2007neural,glaze2015normative}, and object size~\cite{stottinger2006dissociating}. These belief-related biases are well described by Bayesian models of perception, which characterize beliefs as ``priors'' and visual information as ``evidence''. Recent research has explored the influence of people's prior beliefs on their interpretations of information presented in scatter plots. Xiong and colleagues found that people's prior beliefs bias their estimation of the strength of correlations between factors they deem important (e.g., correlation between crime and the availability of guns). In these circumstances, people tend to perceive trends as being stronger or weaker than the ground truth depending on whether the information provided confirms or goes against their beliefs (respectively)~\cite{xiong2022seeing}.  

The current article expands on this research by exploring the conditions under which prior beliefs exert their strongest influence. Research in cognitive psychology finds that prior beliefs are most pervasive under conditions of uncertainty, when the visual information (the evidence) is ambiguous and more difficult to interpret~\cite{gold2007neural, glaze2015normative, glaze2018bias, filipowicz2020pupil, murphy2021adaptive, krishnamurthy2017arousal}. In the context of scatter plots, this work thus proposes that cognitive biases should exert their strongest influence when people have more difficulty perceiving trends in a plot. Previous work has shown that accuracy in perceiving scatter plot trends decreases as variance and uncertainty in the distributions increases~\cite{correll2017regression, rensink2010perception, harrison2014corrweber, wang2022makes}. People also have more difficulty assessing the accuracy of data summaries (e.g., trend lines) when the variance of data points increases~\cite{reimann2021visual}. Therefore, a Bayesian perceptual framework predicts that people's prior beliefs should exert their strongest influence on interpretations of a scatter plot trend when the variance between points is high and people have more difficulty perceiving the trend. Experiments 3 and 4 examine the influence of prior beliefs on trend estimations in scatter plots. Experiment 3 examines whether the influence of prior beliefs depends on the variance (uncertainty) of the data points whereas Experiment 4 examines whether properties of the data (outliers) modulate the influence prior beliefs exert on trend estimations.


\section{General Methods}

\subsection{Participants}
All participants were recruited online from either Amazon Mechanical Turk (MTurk) or Prolific~\cite{palan2018prolific}. On both platforms, participants where only allowed to participate if they lived in the USA, spoke English fluently, were over the age of 18, and had an HIT (for MTurk) or task (for Prolific) approval rating $\geq 95\%$. On MTurk, we recruited participants with at least 1000 completed HITs; on Prolific, we recruited participants with at least 500 completed tasks. For Experiments 2--4, we additionally excluded any participant who had participated in one of our previous experiments. We recruited participants from both Prolific and MTurk in Experiment 1 to compare the quality of responses between platforms. We found no performance differences between platforms on Experiment 1, and so for Experiments 2--4 we recruited exclusively from MTurk as this platform offered a larger pool of eligible participants.

Each study lasted 14 minutes on average (SD = 7 minutes) and participants were paid \$3 USD for participation and up to an additional \$1 USD performance bonus based on response accuracy. In all studies, we excluded participants with either incomplete data (i.e., any missing correlation estimates) or chance-level performance (e.g., their performance was within the 95\% confidence interval of an agent that chose correlation values at random). The study was reviewed and approved by the Western Institutional Review Board (project ID\@: 44793465) and all participants provided informed consent before participating.

\subsection{General correlation estimation task}
In all studies, participants rated the strength of correlations presented in scatter plots. On each trial, participants observed scatter plots with 50 individual points. Depending on the study, 48 or 50 of the individual points were randomly generated using the Kaiser-Dickman (KD) algorithm~\cite{kaiser1962sample} with pre-specified Pearson correlation values that we call the `generative correlation'. The range of generative correlation values used varied between different experiments and are highlighted in each experimental section. Each scatter plot was displayed using \verb|Chart.js| and embedded as an iFrame element within a Qualtrics survey.

We note that participants were only shown positive correlations across all experiments (e.g., generative correlations that could range between 0 and 1). This was a deliberate choice so as to make instructions and training as simple as possible for participants who may not have know about correlations or metrics to quantify a correlation’s strength prior to this study. Additionally, previous research shows that positive correlations are easier for people to interpret (including people with limited statistical training~\cite{kalish2007iterated, harrison2014corrweber}).

At the start of each study, participants completed a tutorial that introduced the concept of correlations and the numerical values used to rate correlation strengths (all with visual examples). Participants then had to successfully rate the strength of three different correlations (0.1, 0.5, and 0.8), with feedback, to move on to the main experiment.

In the main experiment participants used a slider to indicate their estimate of the strength of the correlation in the scatter plots they observed on each trial. The slider ranged between 0.0 and 1.0 (with labels above the slider at every 0.1 increment) and allowed responses in increments of 0.01. To avoid anchoring effects, the slider handle was hidden from participants on each trial until they clicked on the slider to start their response.

At the end of each study, participants completed questionnaires that varied between studies as well as a general demographics questionnaire.

\subsection{General analysis methods}

All four experiments were conducted in two stages: an initial experimental stage in which half of the reported participants completed the experiment, then a replication stage in which the same number of participants completed the same study and the same analyses were used to interpret the results. Analyses were complete three times for each experiment: once after the initial experimental stage, once after the replication, then once on the aggregate data from both the initial and replication stages. All of the aggregate results reported below consist of planned analyses conducted unless otherwise indicated and all reported effects were found in both the initial and replication stages. We used two-tailed tests for all statistics with significance set to $\alpha<0.05$. 

The number of participants per study varied according to the number of possible conditions. Pilot studies found that the accuracy of online participant correlation estimates tended to drop after 25 trials. We thus limited the number of trials each participant completed to no more than 25. Experiment 1 contained 20 different trials and thus all participants provided a response on each trial type. For Experiments 2--4, the number of total conditions exceeded 20 and thus each participant saw a random subset of all possible trials. Our sample size for each experiment was determined to ensure that we collected at least 50 responses across all subjects for each question type.

In a number of the sections below we use linear mixed-effects models (LME), computed using the \verb|lme4|~\cite{bates2014fitting} library in \verb|R|~\cite{rcore}. In all such cases, random intercepts were added for subjects. In cases where aggregate trends were compared between different conditions (e.g., performance on higher versus lower generative correlation values), random intercepts were additionally added for different generative correlation values. In no case did we include random slopes as none of our models converged when these were included (likely due to insufficient trials per subject). All LME degrees of freedom and \textit{p}-values were calculated using the Satterthwaite method implemented by the \verb|lmerTest| library in \verb|R|~\cite{kuznetsova2017lmertest}.
All data, code, and supplemental information about the stimuli used in the creation of this article can be found at \url{https://osf.io/76kp9/}.


\section{Experiment 1: Influence of outliers on interpretations of scatter plots}

Our first experiment examines the influence of outliers on people's perceptions of correlation trends. We aimed to conceptually replicate four main results from Correll and Heer~\cite{correll2017regression}:

\begin{enumerate}
    \item H1: People accurately perceive linear trends of different strengths.
    \item H2: Accuracy decreases for trends with higher variance.
    \item H3: Outliers skew trend perception in the direction of the outliers.
    \item H4: Outliers are given less weight than other points.
\end{enumerate}

One challenge with the method used by Correll and Heer to elicit responses is that people rated their trend estimates using a linear regression line~\cite{correll2017regression, harrison2014corrweber}. Although intuitive, trend lines are salient visual cues~\cite{wang2022makes} and could influence people's interpretation of graph trends. Experiment 1 sought to test the four main hypotheses above using numerical estimates rather than relying on additional annotation methods to elicit responses.

\subsection{Participants and methods}

We recruited 102 participants for Experiment 1 (48 from MTurk, 54 from Prolific). Data from 9 participants were discarded before analysis because data were missing or they had chance-level performance (criteria outlined in the general methods), leaving a total of 93 participants (47 identifying as women, 46 as men; mean age 41 years).

\begin{figure*}
  \centering
  \begin{subfigure}[c]{0.22\textwidth}
    \includegraphics[width=\textwidth]{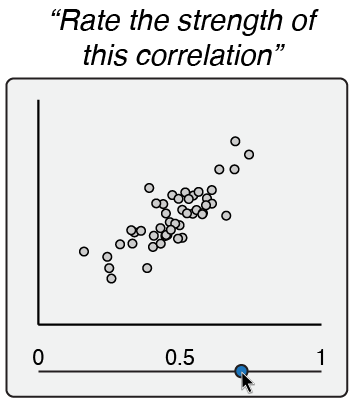}
    \caption{}\label{fig:1a}
  \end{subfigure}
  \hspace{1pc}
  \begin{subfigure}[c]{0.3\textwidth}
    \includegraphics[width=\textwidth]{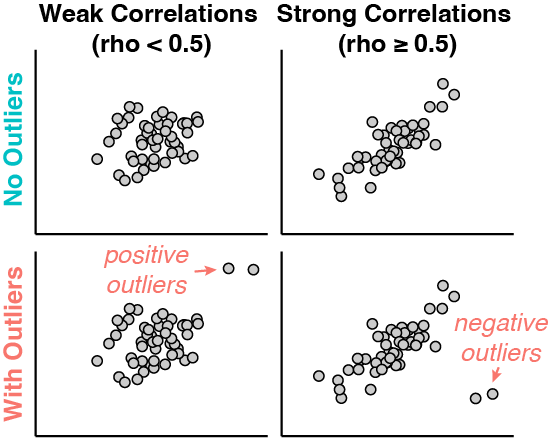}
    \caption{}\label{fig:1b}
  \end{subfigure}
  \hspace{1pc}
  \begin{subfigure}[c]{0.3\textwidth}
    \includegraphics[width=\textwidth]{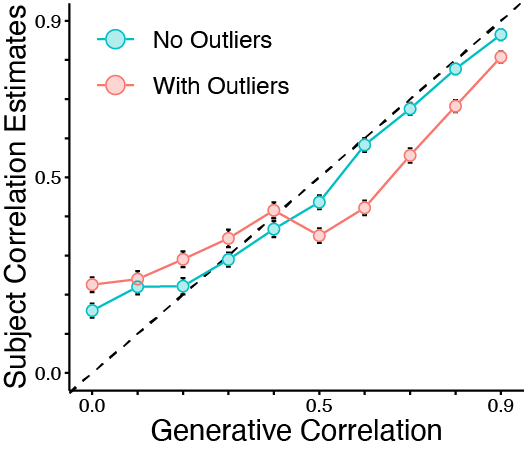}     
    \caption{}\label{fig:1c}
  \end{subfigure}
  \vspace{1pc}
  \begin{subfigure}[d]{0.5\textwidth}
    \includegraphics[width=\textwidth]{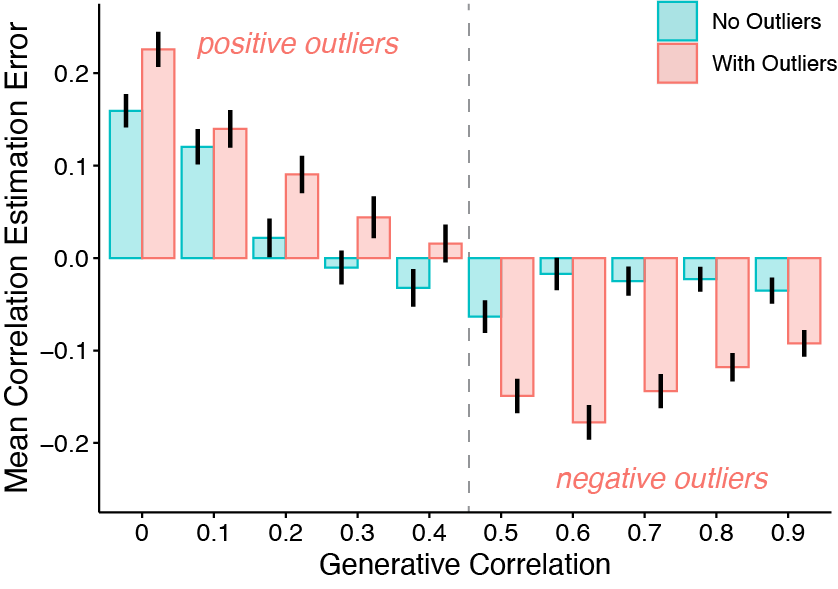}
    \caption{}\label{fig:1d}
  \end{subfigure}
  \hspace{1pc}
  \begin{subfigure}[e]{0.35\textwidth}
    \includegraphics[width=\textwidth]{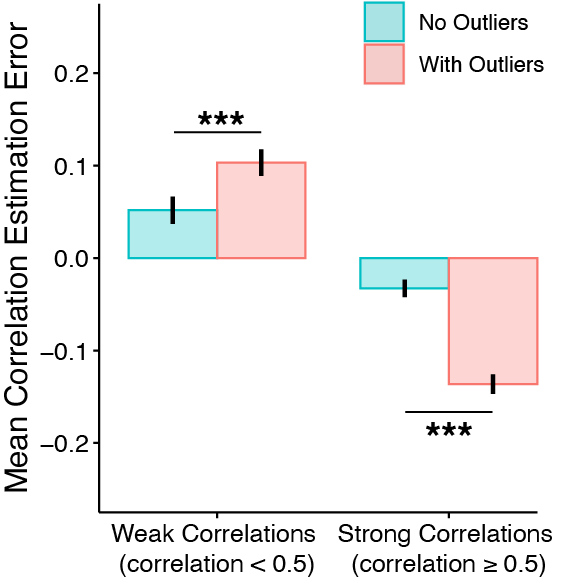}
    \caption{}\label{fig:1e}
  \end{subfigure}
  \caption{\label{fig:fig1} Influence of outliers on correlation
    estimates. (a) Example trial. On each trial participants used a
    slider to rate the strength of a correlation presented on a
    scatter plot. (b) Correlations were presented either without
    outliers (teal) or with outliers (pink). Outliers for weaker
    correlations (correlations $<0.5$) increased the Pearson
    correlation value of the other dots (positive outliers) whereas
    outliers for stronger correlations (correlations $\geq 0.5$)
    decreased the correlation of the other dots (negative
    outliers). (c) Participant correlation estimates closely tracked
    the intended correlations, but participants tended to
    over-estimate correlation with positive outliers, and
    underestimate correlations with negative outliers. These
    differences are also reflected in the correlation estimation
    errors across (d) all generative correlation values and (e) mean
    errors for weak and strong correlations. Points and bars in panels
    (c--e) correspond to means and error bars indicate one standard
    error of the mean. ***: $p<.001$}
\end{figure*}

Participants rated the correlation strength of 20 scatter plots: 10 scatter plots showing correlations without any outliers, and 10 plots with outliers. The order in which the plots were presented was randomized between participants. The graph title instructed participants to ``Rate the strength of the correlation'' (Fig.~\ref{fig:fig1}a).

Scatter plots without outliers contained 50 points generated using the KD algorithm. The generative correlation values used for the algorithm ranged from 0.0 to 0.9 in increments of 0.1 and participants saw one scatter plot for each correlation value (10 total). For this experiment, we did not include trials with generative correlation values of 1.0 (i.e., perfect correlations) because we did not expect much variability in participant responses (as these values are easy to estimate). However, we note that we subsequently added trials with generative correlations of 1.0 to Experiment 2. 

Scatter plots with outliers consisted of 48 points randomly generated using the KD algorithm and two additional outlier points that either increased or decreased the generative correlation value. The type of outlier depended on the generative correlation value. Outliers that increased the correlation strength (``positive'' outliers) exert an influence when correlations are relatively weak. Conversely, outliers that decrease correlation strength (``negative'' outliers) exert an influence primarily when correlation strength is relatively strong. As such, we used positive outliers on trials with relatively weak generative correlations (i.e., correlations $<0.5$) and negative outliers for relatively strong generative correlations (i.e., correlation $\geq0.5$; Fig.~\ref{fig:fig1}b).

We generated outliers such that they were clearly distinct from the 48 other points on the plot but also generated with random jitter so that the same outliers would not be seen participant more than once throughout the task. With these general criteria we used the following steps to generate the horizontal (x) and vertical (y) outliers positions:

\begin{enumerate}
\item Each outlier's position x-axis position was generated as a random sample from a uniform distribution between 1.5 times the horizontal mean of the 48 points and the maximum range we allowed for the horizontal position of the points.
\item For positive outliers, the y-axis position was generated as a random sample between the vertical mean of the 48 points plus 3 and 4 standard deviations from the vertical mean.
\item Negative outlier positions and the y-axis were generated using the same calculation in step 2 with an additional step of subtracting a scalar value from the y-position to make the influence of the outliers comparable to the influence of the positive outliers---that is, the net increase the outliers added to the Pearson correlation for the lower correlation values {0.0, 0.1, 0.2, 0.3, 0.4} was mirrored by a similar decrease for higher correlation values {0.5, 0.6, 0.7, 0.8, 0.9} respectively.
\end{enumerate}

On average, outliers increased (for positive outliers) or decreased (for negative outliers) the Pearson correlation values of the scatter plots by 0.28 (standard deviation = 0.05; examples of outliers can be found in Fig.~\ref{fig:fig1}b). 

\subsection{Participants can accurately represent different correlation values but accuracy depends on plot variance}

Overall, on trials without outliers, participant correlations estimates closely matched the generative correlations they were presented (LME measuring subject correlation estimates vs generative correlations: $\beta = .82, t(1,764) = 42.48, p<.001$; Fig.~\ref{fig:fig1}c). However, accuracy depended on correlation value---absolute correlation estimation errors tended to be higher for lower generative correlation values (LME measuring subject absolute correlation estimation error vs generative correlations: $\beta = -.84, t(1,764) = -9.32, p<.001$). Correlation values also serve as measures of variance, with lower values reflecting less coherence between trends on both axes. Therefore, these two results support H1 and H2 in that people are able to accurately estimate correlation trends, but are more precise for higher correlation values, which are more coherent and less variable.

\subsection{Participants are highly susceptible to outliers but weight these less than other points}

On trials where outliers were present, participant correlation estimates deviated farther from the generative correlation (LME measuring interaction between generative correlation value and presence of outliers on participant correlation estimates: $\beta =-.22, t(1,764) = -8.11, p<.001$). On average, participants overestimated weak generative correlation when positive outliers were displayed (LME testing main effect of outliers on estimation error for correlations \textless 0.5: $\beta = .05, t(832)=5.04, p<.001$) and underestimated strong generative correlations when negative outliers were displayed (LME testing main effect of outliers on estimation error for correlations \textgreater 0.4: $\beta=-.10, t(832)=-11.23, p<.001$). Negative outliers also had a slightly larger effect than positive outliers (LME testing interaction between outliers being present {present, not present} and outlier type {positive, negative} on absolute correlation estimation errors: interaction $\beta = -.03, t(1756)=-2.29, p=.022$). Thus, we find support for H3 in that participant responses were overall influenced by outliers: their estimates increased and decreased when outliers increased and decreased, respectively, the correlation of the scatter plot. 

\begin{figure*}[t]
  \centering
  \begin{tabular}{cc}
        \includegraphics[width=0.8\columnwidth]{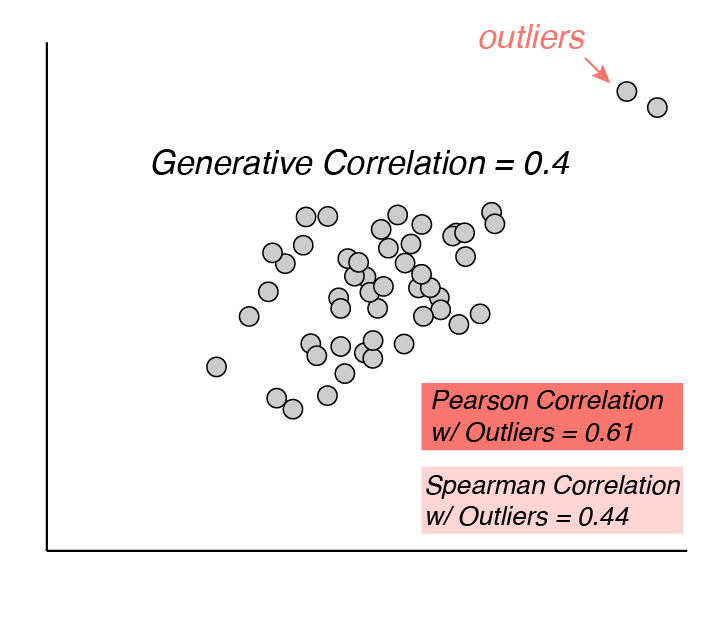} & \includegraphics[width=0.85\columnwidth]{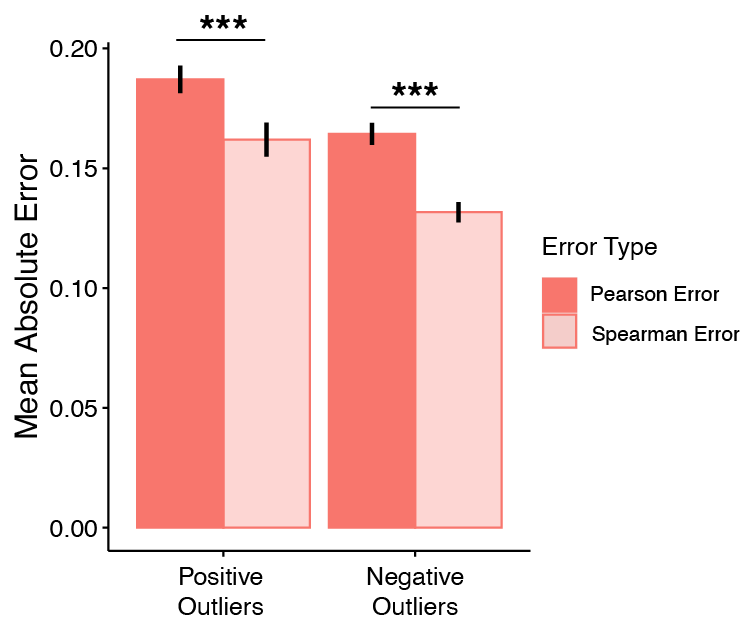} \\
        (a)\label{fig2a} & (b)\label{fig2b} \\
    \end{tabular}
    \caption{\label{fig:fig2} Participants down-weigh outliers compared to other points. (a) Example of the influence of outliers on a plot with 48 points generated with a generative correlation value of 0.4 and two added positive outliers. Pearson correlations are computed by equally weighing the magnitude of the position of all of the points. This makes Pearson correlations particularly susceptible to outliers (e.g., increasing correlation in the example from 0.40 to 0.61). Conversely, Spearman correlation compare the rank of each point relative to the others, ignoring the magnitude of the difference between points. As a result, outliers have a smaller effect on Spearman correlation coefficients (e.g., only increasing the correlation in the example from 0.40 to 0.44). (b) Participant correlation estimates were closer to plot Spearman correlation values than Pearson correlation values (smaller mean absolute error) for plots with either positive or negative outliers.***: $p<0.001$. Bars in (b) correspond to means and error bars correspond to one standard error of the mean; ***: $p<0.001$.}
\end{figure*}

Although outliers influenced people’s correlation estimates, they did not have as strong an influence as would be expected if people were weighing outliers to the same extent as other points. We measured the influence of outliers by comparing two possible ways in which outliers could influence correlation estimates (examples given in Fig.~\ref{fig:fig2}a):

\begin{enumerate}
\item \textit{People give equal weight to each point in the scatter plot when estimating the correlation:} A process like this would be akin to a Pearson correlation, which takes into account the magnitude of the relationship between all of the points in the scatter plot. This type of processing predicts a relatively large influence of outliers.
\item \textit{People focus primarily on points that are consistent in the scatter plot and down-weigh outliers:} A process like this would be akin to a Spearman correlation, where the rank between points is used to estimate correlation strength instead of magnitude. This type of processing predicts a relatively small influence of outliers.
\end{enumerate}

To compare these possibilities, we measured the extent to which participant correlation estimates differed from the Pearson and Spearman correlation values computed from the points participants observed in each trial. Overall, participant estimates were closer to Spearman estimates than Pearson estimates on trials with either positive or negative outliers (LME testing main effect of error type on absolute correlation estimation error---correlations \textless 0.5 with positive outliers: $\beta = -.05, t(832)=-6.55, p<.001$; correlations \textgreater 0.4 with negative outliers: $\beta = -.07, t(832)=-9.91, p<.001$; Fig.~\ref{fig:fig2}b). These results support H4 as they indicate that participants give less weight to outliers than they do other points in the correlation plots.

Taken together, our results support the four hypotheses suggested from previous research: (1) people can accurately represent and report correlations using our new paradigm, (2) accuracy is lower for plots with higher variance (lower correlation values), (3) outliers influence participant estimates, but (4) people are still somewhat robust to the influence of outliers.

We next examined whether another common scatter plot feature, trend lines, could be used as a potential tool to mitigate the influence of outliers on people's perception of correlation strength.


\section{Experiment 2: Influence of trend lines on interpretations of scatter plots}

Trend lines are another common visual feature added to scatter plot to highlight the direction and strength of a correlation. These are generally computed using ordinary least squares (OLS) calculations to provide a ``line of best fit'' across all of the points on the plot. Trend lines are highly salient annotation markers that can be useful for highlighting trends~\cite{wang2022makes}. However, if trend lines are computed using OLS methods, these can also be affected by outliers, similar to the way outliers affect traditional linear statistics.

Experiment 2 examines the extent to which trend lines influence people's interpretation of correlation estimates and how these interact with the presence of outliers. We tested three main hypotheses:

\begin{enumerate}
    \item H1: When no outliers are present, trend lines improve trend estimation accuracy.
    \item H2: When outliers are present, trend lines that include outliers exacerbate the influence of outliers on trend estimations.
    \item H3: When outliers are present, trend lines that omit outliers reduce their influence on trend estimations.
\end{enumerate}

\subsection{Participants and methods}

A total of 421 participants were recruited for Experiment 2 from MTurk. Data from 41 participants were discarded before analysis due to incomplete responses or chance-level performance, leaving a total of 380 participants (145 identifying as women, 228 as men, 2 as transgender, 1 as non-conforming, 1 as another gender, and 3 preferred not to answer; mean age 38 years).

Participants rated the strength of correlations in 22 scatter plots that ranged between 0.0 and 1.0 and could be presented with or without outliers. However, in all cases there was an equal likelihood that a trend line would or would not appear on each scatter plot, such that participants were not exposed to all possible combinations of generative correlation values, outliers, and trend lines. On trials that did not contain outliers, the trend lines were computed as the OLS line over all 50 points. On trials that included outliers, the trend lines were computed one of two ways: (1) using all points (i.e., including outliers), or (2) excluding outliers (i.e., only based on the first 48 generate points; Fig.~\ref{fig:fig3}a). The order in which the scatter plots were presented was randomized between participants.

\subsection{Correlations are perceived as stronger when presented with trend lines}

\begin{figure*}[t]
    \centering
    \begin{minipage}{.35\textwidth}
    \begin{subfigure}[t]{\textwidth}
    \includegraphics[width=\textwidth]{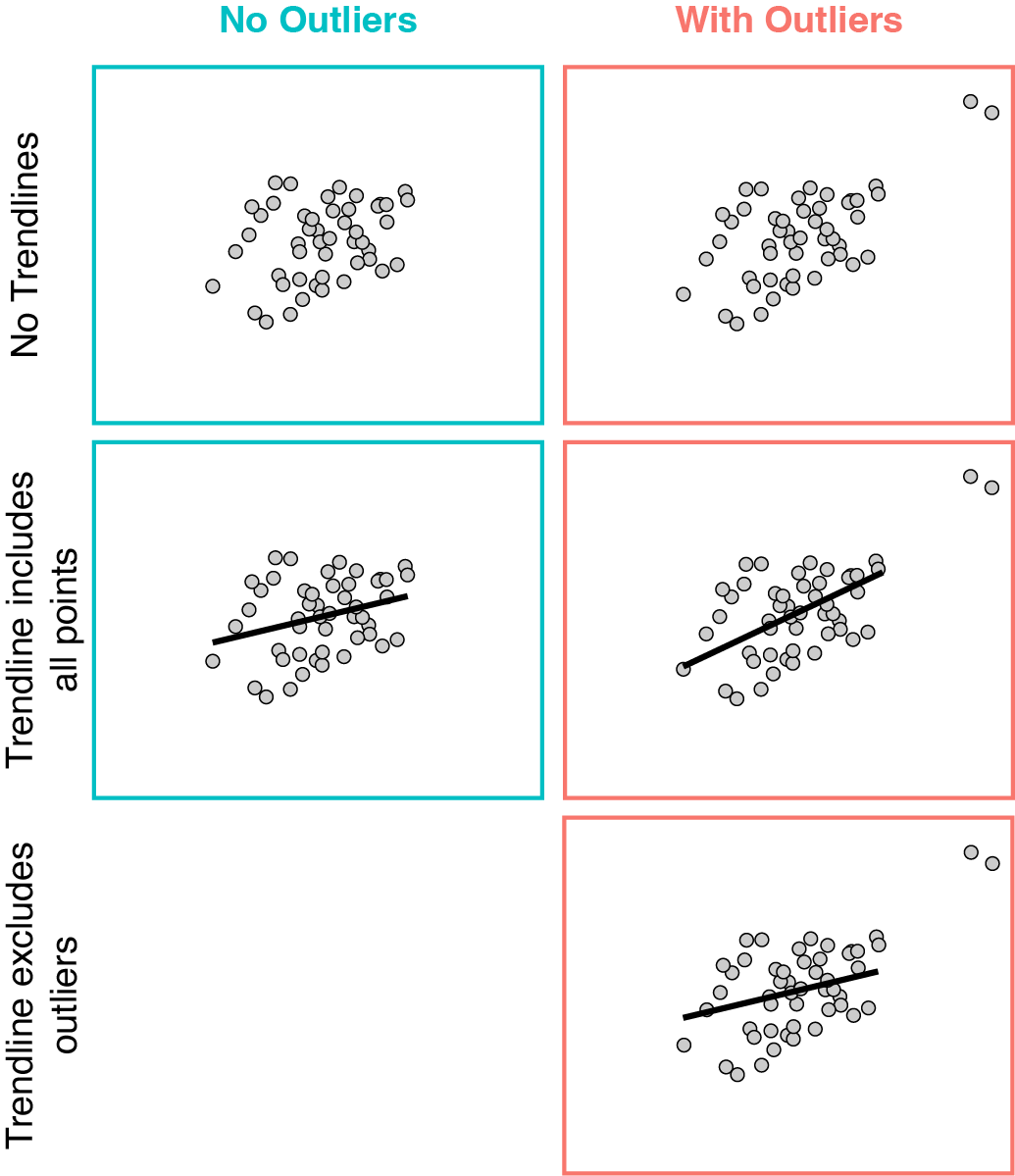}  
    \caption{\label{fig:3a}}
    \end{subfigure}
    \end{minipage}
    \hspace{1pc}
    \begin{minipage}{.34\textwidth}
    \begin{subfigure}[t]{\textwidth}
    \includegraphics[width=\textwidth]{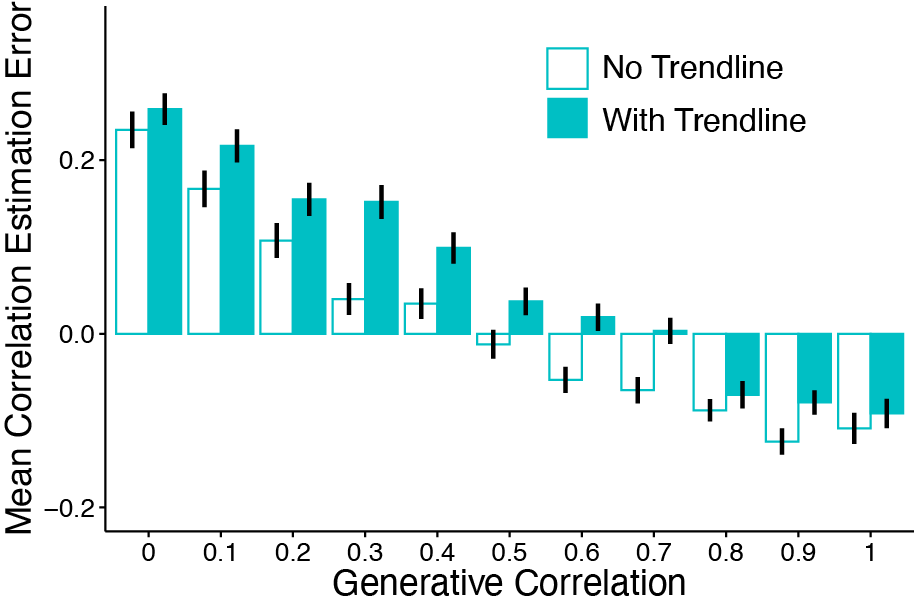}
    \caption{\label{fig:3b}}
    \vspace{1pc}
    \includegraphics[width=\textwidth]{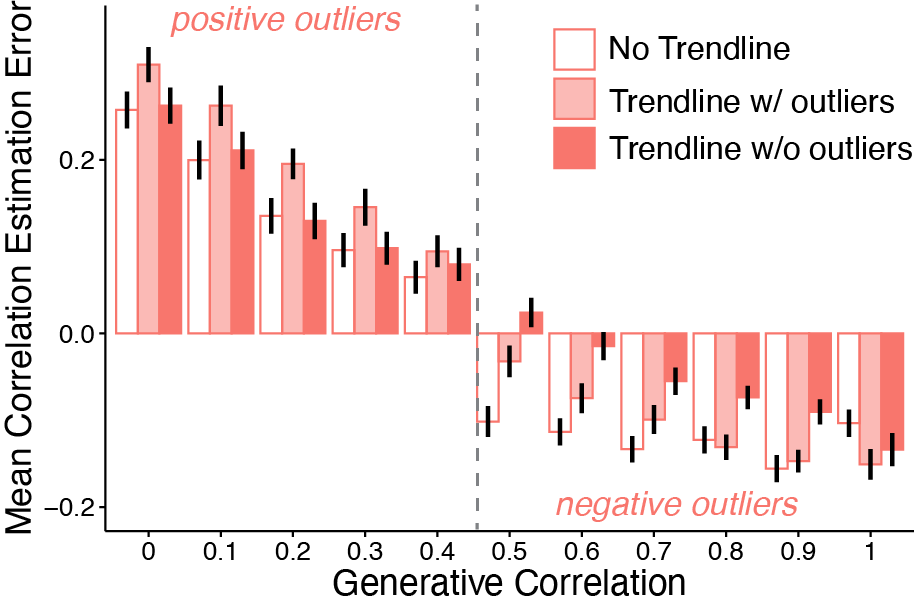}
    \caption{\label{fig:3c}}
    \end{subfigure}
    \end{minipage}
    \hspace{1pc}
    \begin{minipage}{.2\textwidth}
    \begin{subfigure}[t]{\textwidth}
    \includegraphics[width=\textwidth]{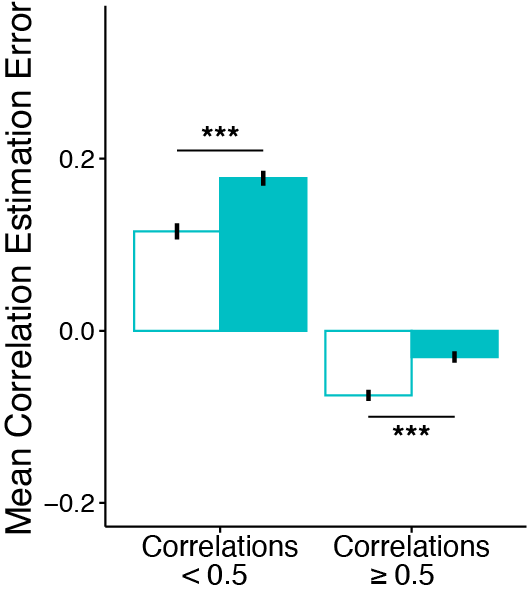}
    \caption{\label{fig:3d}}
    \vspace{1pc}
    \includegraphics[width=\textwidth]{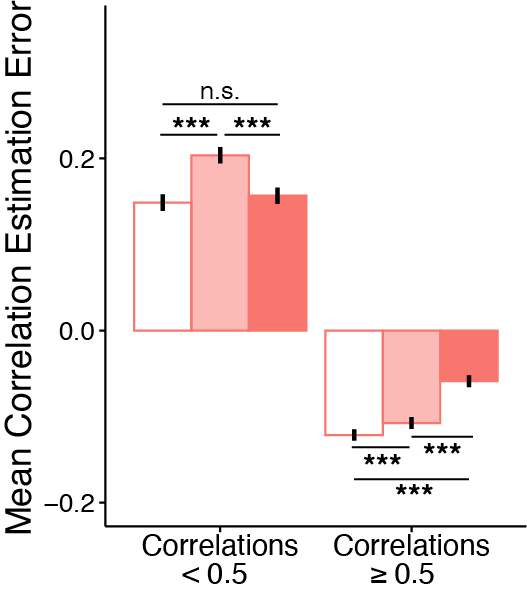}
    \caption{\label{fig:3e}}
    \end{subfigure}
    \end{minipage}
    \caption{\label{fig:fig3} Influence of trend lines on correlation
      estimations. (a) Participants rated the strength of correlations
      that were presented with or without trend lines and with or
      without outliers. When trend lines were presented on trials with
      outliers, the trend lines were computed either by including or
      excluding the outliers. (b,d) On trials without outliers,
      participants rated correlations with trend lines as being
      overall stronger than those presented without trend lines. (c,e)
      Participants rated correlations with trend lines that excluded
      outliers as being closer to the generative correlations compared
      to correlations with trend lines included outliers. Moreover, on
      trials with strong correlations and negative outliers,
      participants rated correlations with trend lines that excluded
      outliers as closer to the generative correlations than when no
      trend lines were displayed. This was not true for trials with
      weak correlations and positive outliers. Bars in panels (b--e)
      correspond to means and error bars correspond to one standard
      error of the mean; ***: $p<.001$.}
\end{figure*}

Overall, on trials with no outliers, participants rated correlations as stronger when trend lines were displayed than when they were not (Fig.~\ref{fig:fig3}b). This was true for both weaker and stronger generative correlation values (LME comparing estimation error on trials with versus without trend lines---correlations $<0.5$: $\beta=.53, t(1,106)=5.71, p<.001$; correlations $\geq 0.5$: $\beta=.46, t(1,449)=5.60, p<.001$; Fig.~\ref{fig:fig3}d). Thus, contrary to H1, rather than provide an overall improvement to participant accuracy, trend lines had the effect of making participants perceive correlations as stronger than they actually were.

\subsection{Trend lines that exclude outliers can mitigate the effect of outliers}

We next examined whether excluding outliers from the calculation of a trend line mitigated the influence of outliers on participant correlation estimates. When rating scatter plots with weaker correlation (i.e., correlations $< 0.5$ with positive outliers), participant correlation estimations were closer to the generative correlation when trend lines did not include outliers compared to trend lines that did include outliers, but these estimates were no closer to the generative correlation than estimates without outliers (LME contrast comparing correlation estimation errors---trials with trend lines that excluded outliers versus trend lines that included outliers: $\beta=.52, t(1,733)=5.30, p<.001$; trials with trend lines that excluded outliers versus trials with no trend lines: $\beta=-.01, t(1,730)=-.99, p=.324$). 

In contrast, when rating scatter plots with stronger correlations (i.e., correlations $\geq 0.5$ with negative outliers), participant estimates were closer to the generative correlation when trend lines excluded outliers than either scatter plots with trend lines that included outliers or scatter plots with no trendline (LME contrast comparing correlation estimation errors---trials with trend lines that excuded outliers versus trend lines that included outliers: $\beta=-.55, t(2,269)=-6.55, p<.001$; trials with trend lines that excluded outliers versus trials with no trend lines: $\beta=-.69, t(2,296)=-8.08, p<.001$; Fig.~\ref{fig:fig3}d,e). 

We therefore find support for H2 in that participant estimates were more influenced by outliers when these were included in the calculation of the trend lines. We additionally find partial support for H3 in that trend lines that did not include outliers mitigated the effect of negative, but not positive outliers.

Taken together, the results Experiments 1 and 2 demonstrate that visual aspects and features of scatter plots can influence people's interpretation of correlations. We next examined the extent to which people's beliefs influence and interact with visual aspects of scatter plots to influence perceptions of correlation trends. 


\section{Experiment 3: Influence of cognitive biases on interpretations of scatter plots}

Some recent prior work demonstrates that people beliefs can bias and distort their interpretation of scatter plots that convey information related to their beliefs~\cite{xiong2022seeing}. Research in cognitive psychology also demonstrates that people's beliefs can bias there perception of patterns and objects~\cite{jabar2017tuned,gold2007neural,glaze2015normative,stottinger2006dissociating}. This research predicts that beliefs exert their strongest influence under conditions of uncertainty, when the visual information is ambiguous and more difficult to interpret~\cite{gold2007neural,glaze2015normative,glaze2018bias,filipowicz2020pupil,murphy2021adaptive,krishnamurthy2017arousal}. Although previous work has demonstrated that beliefs do play a role in the perception of scatter plots, it is unclear whether these effects are more prominent under conditions of uncertainty.

Experiment 3 examines if and when people's beliefs influence their perception of correlations in scatter plots. Correlations are inherently measures of uncertainty, with higher values indicating more consistency between two variables and lower values indicating more uncertainty. Experiment 3 thus sought to test two main hypotheses:

\begin{enumerate}
    \item H1: People's estimates are biased in the direction of their beliefs.
    \item H2: The influence of beliefs are strongest for weaker correlations.
\end{enumerate}

\begin{figure*}[h!]
    \centering
    \begin{subfigure}[t]{0.4\textwidth}
        \includegraphics[width=\textwidth]{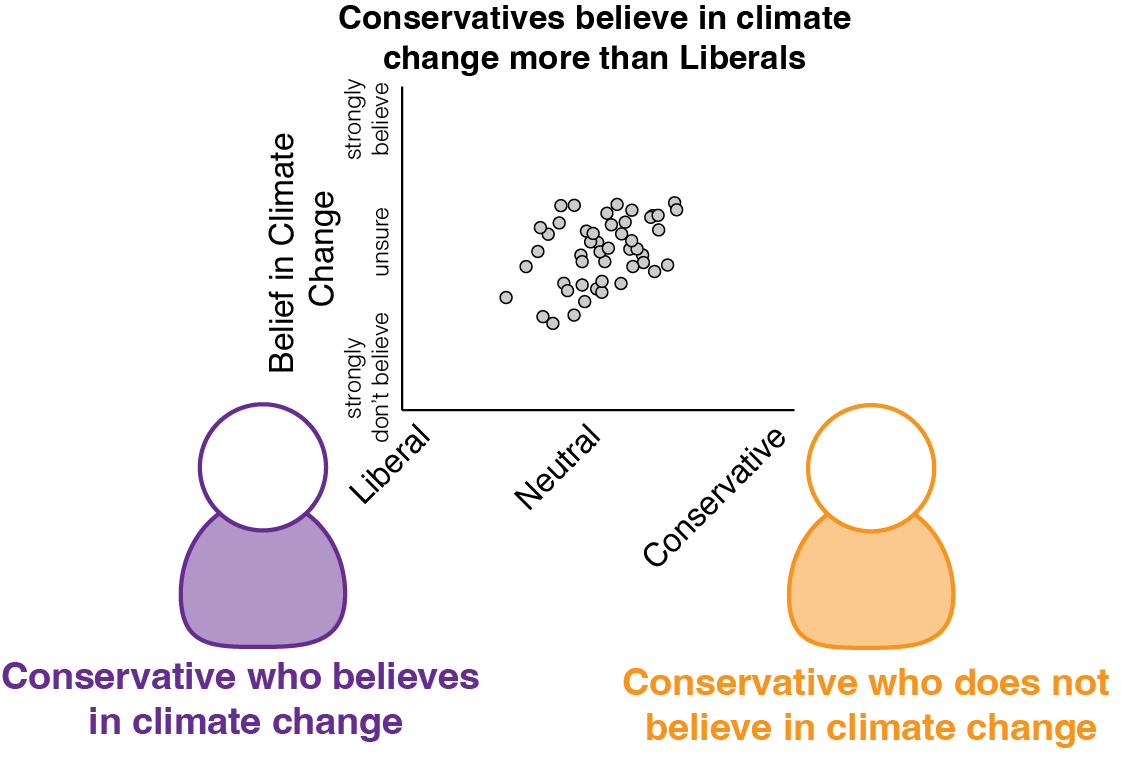}
        \caption{\label{fig4a}}
    \end{subfigure}
    \hspace{1pc}
    \begin{subfigure}[t]{0.25\textwidth}
        \includegraphics[width=\textwidth]{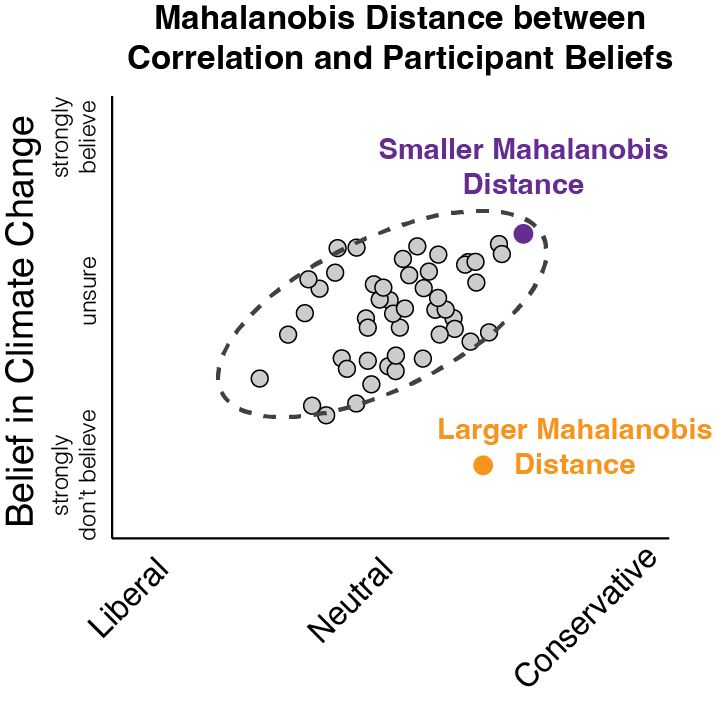}
        \caption{\label{fig4b}}
    \end{subfigure}
    \vspace{1pc}
    \begin{subfigure}[t]{0.7\textwidth}
        \includegraphics[width=\textwidth]{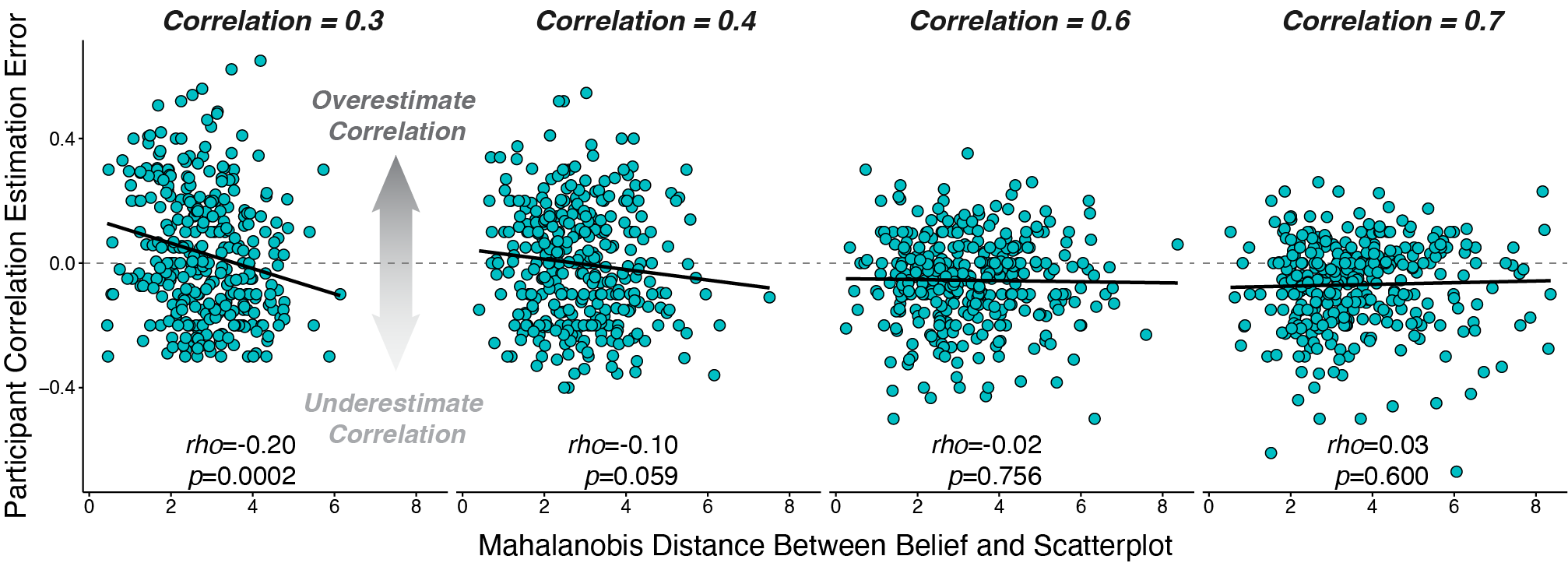}
        \caption{\label{fig4c}} 
    \end{subfigure}
    \caption{\label{fig:fig4} Influence of beliefs on correlation estimations without outliers. (a) Participants rated the strength of correlations in scatter plots with labels and chart titles showing a trend between a belief (on the y-axis) and a trait (on the x-axis). Participant beliefs and traits related to each scatter plot were measured via post-study questionnaires. These scores were used to measure a distance between people's beliefs/personality and the trends they observed. (b) Mahalanobis distance was used to measure distance between people's beliefs/traits and the scatter plot trends they observed. This measure captures the distance between a point (where the person falls in belief/personality space) and a distribution (the points in the scatter plot), taking into account the distribution's covariance (dashed ellipse). (c) On trials with weak generative correlations (correlation = 0.3), participants tended to overestimate the strength or trends that were close to their beliefs (i.e., confirmed their beliefs; low Mahalanobis distance) and underestimate the strength of trends that were far from their beliefs (i.e., went against their beliefs; high Mahalanobis distance). Distance from participant beliefs had no effect when generative correlations were 0.4 or above. Trend lines correspond to ordinary least-squares regression lines. All reported statistics correspond to Spearman correlation values.}
\end{figure*}

\subsection{Participants and methods}
A total of 407 participants were recruited from MTurk for Experiment 3. Data from 43 participants were discarded before analysis due to incomplete responses or chance-level performance, leaving a final sample of 364 participants (127 identifying as women, 233 as men, 1 as transgender, 1 as non-conforming, 1 as another gender, and 1 preferred not to answer; mean age 37 years).

Participants completed two sets of correlation estimations. In the first set, participants rated the correlation strength of 11 scatter plots with correlations from 0.0 to 1.0 in increments of 0.1. The instructions given were identical to those in Experiment 1 and the order of the scatter plots were randomized between participants.

In the second set, participants rated the strength of hypothetical correlations linking people's beliefs and different traits. After some examples, participants rated the strength of 10 additional correlations taken from a set that showed hypothetical correlations between a labelled trait on the x-axis (e.g., “Political Preference”) and a labelled belief on the y-axis (e.g., “Belief in Climate Change”; Fig.~\ref{fig:fig4}a, Tables~\ref{tab:tab1} and ~\ref{tab:tab2}). The correlation strength could be 0.3, 0.4, 0.6, or 0.7, values which allowed some range in responses above or below the true correlation value. A title was added to the top of each scatter plot that summarized the trend displayed in the plot (e.g., “Conservatives believe in climate change more than Liberals” above a plot showing a positive correlation between degree of conservativeness and belief in climate change; Fig.~\ref{fig:fig4}a). Each axis also contained label-specific sets of sub-labels to anchor the correlations (e.g., “strongly liberal” and “strongly conservative” on the left and right side, respectively, of a “Political Preference” x-axis; Tables~\ref{tab:tab1} and ~\ref{tab:tab2}). The direction of the correlations were randomized between participants such that some participants saw one direction between traits and beliefs (e.g., “Conservatives believe more in climate change than liberals”) and others saw the opposite trend (e.g., “Liberals believe in climate change more than conservatives”; see tables S1--S4 in the supplementary material for a full list of graph titles used in our study).

The following traits were included in this study: political preference, open-mindedness, prosociality, and age. We included the following beliefs in this study: belief in climate change, attitudes towards electric vehicles, attitudes towards renewable energy, and trust in science.

Participants then completed a series of questionnaires that related to the combinations of beliefs and traits they had been exposed to throughout the study. For the traits, we used the following surveys: Actively Open-Minded Thinking scale to measure open-mindedness~\cite{baron2019actively}, Prosociality Scale to measure prosociality~\cite{caprara2005new}, a single 5-point likert item asking “What are your political views?” with answers ranging from “Strongly Liberal” to “Strongly Conservative” to measure political preferences, and a demographics questionnaire to ask for age. For the beliefs we used the following scales: Attitudes towards Climate Change scale to measure beliefs in climate change~\cite{mildenberger2019beliefs}, Trust in Science and Scientists scale to measure trust in science~\cite{mccright2013influence}, and we created two scales to measure attitudes towards electric vehicles and attitudes towards renewable energy (scales in supplementary material).

We note that participants completed the belief and trait questionnaires \emph{after} estimating the correlations in the labelled scatter plots because we wanted to avoid them potentially anticipating the scatter plots they may observe. However, instead of measuring \emph{prior} beliefs, this ordering may have unintentionally had participants report \emph{posterior} beliefs influenced by the trends they estimated. Xiong and colleagues previously found that the order in which questionnaires are administered does not influence participant's belief responses for plausible, ``belief triggering'' correlations (e.g., correlation between crime and guns~\cite{xiong2022seeing}). To examine whether the labelled correlations influenced the beliefs people reported, we used independent t-tests to compare questionnaire scores from participants who randomly saw correlations with one trend (e.g., ``conservatives believe in climate change more than liberals'') to those who randomly saw correlations with the opposite trend (e.g., ``liberals believe in climate change more than conservatives''). We found no differences in participant belief and trait questionnaire scores when comparing the trends participants saw (all $|t|<1.7$ and all $p>.08$; see Figures S1 and S2 in the supplementary materials for more details), suggesting that participant questionnaire scores were not influenced by the trends participants observed and most likely represented prior rather than posterior beliefs. Additionally, the beliefs we tested were all closer to the plausible beliefs tested by Xiong and colleagues, and so our results support their finding that questionnaire order does not influence the measure of more plausible beliefs~\cite{xiong2022seeing}.

\begin{table*}
\begin{center}
  {\color{changeColorVar}\begin{tabular}{p{2.8cm} p{2.8cm} p{2.8cm} p{2.8cm} p{2.8cm}}
 \textbf{Belief} & \textbf{Axis Label} & \textbf{Sub-label 1} & \textbf{Middle} & \textbf{Sub-label 2} \\
 \toprule
Attitudes towards climate change & Belief in \newline climate change & Strongly believe & Unsure & Strongly \newline do not believe \\  
\midrule
Trust in science & Trust in science & Strongly trust & Unsure & Strongly \newline do not trust \\
\midrule
Attitudes \newline  towards EVs & Preference \newline for EVs & Prefer gas cars & Indifferent \newline about EVs & Prefer EVs \\
\midrule
Attitudes towards renewables & Preference \newline to choose renewable energy & Never & Sometimes & Always \\
  \bottomrule
\end{tabular}}
\caption{\label{tab:tab1} Axis labels and sub-labels on the x-axis for each belief.}
\end{center}
\end{table*}
 
\begin{table*}
\begin{center}
  {\color{changeColorVar}\begin{tabular}{p{2.8cm} p{2.8cm} p{2.8cm} p{2.8cm} p{2.8cm}}
 \textbf{Trait} & \textbf{Axis Label} & \textbf{Sub-label 1} & \textbf{Middle} & \textbf{Sub-label 2} \\
 \toprule
Open-mindedness & Open-mindedness & Very \newline open-minded & Somewhat \newline open-minded & Very stubborn \\  
\midrule
Prosociality & Consideration \newline  for others & Very considerate & Somewhat \newline  considerate & Very selfish \\
\midrule
Political \newline preference & Political \newline preference & Liberal & Neutral & Conservative \\
  \bottomrule
\end{tabular}}
\caption{\label{tab:tab2} Axis labels and sub-labels on the y-axis for each trait.}
\end{center}
 \end{table*}
 
\subsection{Measuring distance of beliefs from scatter plot trend}

We used participant responses on their post-study questionnaires to measure the distance between their beliefs and the trend they observed on each trial. Participant questionnaire scores and scatter plot point positions were both normalized so participant scores could be placed in the scatter plot graphs.

We used Mahalanobis distance to measure the distance between the correlation in the scatter plot and people's beliefs, which measures the distance of a point (or set of points) from a distribution. Mahalanobis distance ($D_m$) is measured as:

\[ D_m = \sqrt{(x-\hat{\mu})S^{-1}(x-\hat{\mu})} \]

where $x$ is a vector containing the coordinates of a point to be compared to a distribution, $\hat{\mu}$ corresponds to the centroids of the correlation distribution along different coordinate axes, and $S^{-1}$corresponds to the covariance matrix of the distribution. In our case, a person’s point is a 2D vector along the belief and trait dimensions and compared to the centroids and covariance of the scatter plot of dots (Fig.~\textbf{}\ref{fig:fig4}b).

We note that Mahalanobis distance is constrained only to take on positive real values and tended to be positively skewed. We therefore used non-parametric Spearman correlations in our analyses comparing belief distance to other participant metrics.

\subsection{Beliefs influence correlation ratings for weak, but not strong, correlation values}

Overall, people's beliefs influenced their perception of weak but not strong correlation trends (Fig.~\ref{fig:fig4}c). On trials where the generative correlation was weakest (i.e., correlation = 0.3) we found a negative correlation between a participant's estimation errors and the distance of their beliefs from the trend displayed in the scatter plot (Spearman correlation [95\% CI]: $\rho=-.20 [-.30,-.10], p<.001$). This indicates that for these weak correlations, participants tended to overestimate the strength of trends that were consistent with their beliefs (i.e., low Mahalanobis distance) and underestimate the strength of trends that went counter to their beliefs (i.e., high Mahalanobis distance; Fig.~\ref{fig:fig4}c). We did not find this influence of beliefs for any of the stronger correlations (Fig.~\ref{fig:fig4}c).

This result provides partial support for H1 and H2, indicating that beliefs can exert a small influence on people's interpretations of trends, but primarily under circumstances where trends in the graphs are weak. When the trends are stronger, the evidence tends to dominate over people's beliefs. We next wanted to examine whether people's beliefs interact with visual aspects of the scatter plot, notably, whether these biases are magnified or mitigated by outliers.


\section{Experiment 4: Interaction between cognitive biases and outliers on interpretations of scatter plots}

Having established that both visual features and beliefs influence people's perceptions of correlations in scatter plots, we conducted a last experiment to examine the extent to which these features interact. We specifically hypothesized that outliers magnify the influence of beliefs on perceptions of correlations.

\subsection{Participants and methods}
A total of 809 participants were recruited from MTurk for Experiment 4. Data from 86 participants were discarded before analysis due to missing data or chance-level performance, leaving a total sample of 723 participants (298 identifying as women, 418 as men, 2 as transgender, 1 as non-conforming, 1 as another gender, and 3 preferred not to answer; mean age 38 years).

The design and analysis of participant data was identical to that used in Experiment 3, except that scatter plots could additionally include outliers (positive outliers for correlations of 0.3 and 0.4 and negative outliers for correlations of 0.6 and 0.7). The addition of outliers doubled the total number of possible conditions. Since we did not increase the number of conditions per participant, we instead doubled the number of participants from Experiment 3. 

\subsection{Outliers do not interact with influence of beliefs}

\begin{figure*}[h!]
    \centering
    \begin{subfigure}[t]{0.7\textwidth}
        \includegraphics[width=\textwidth]{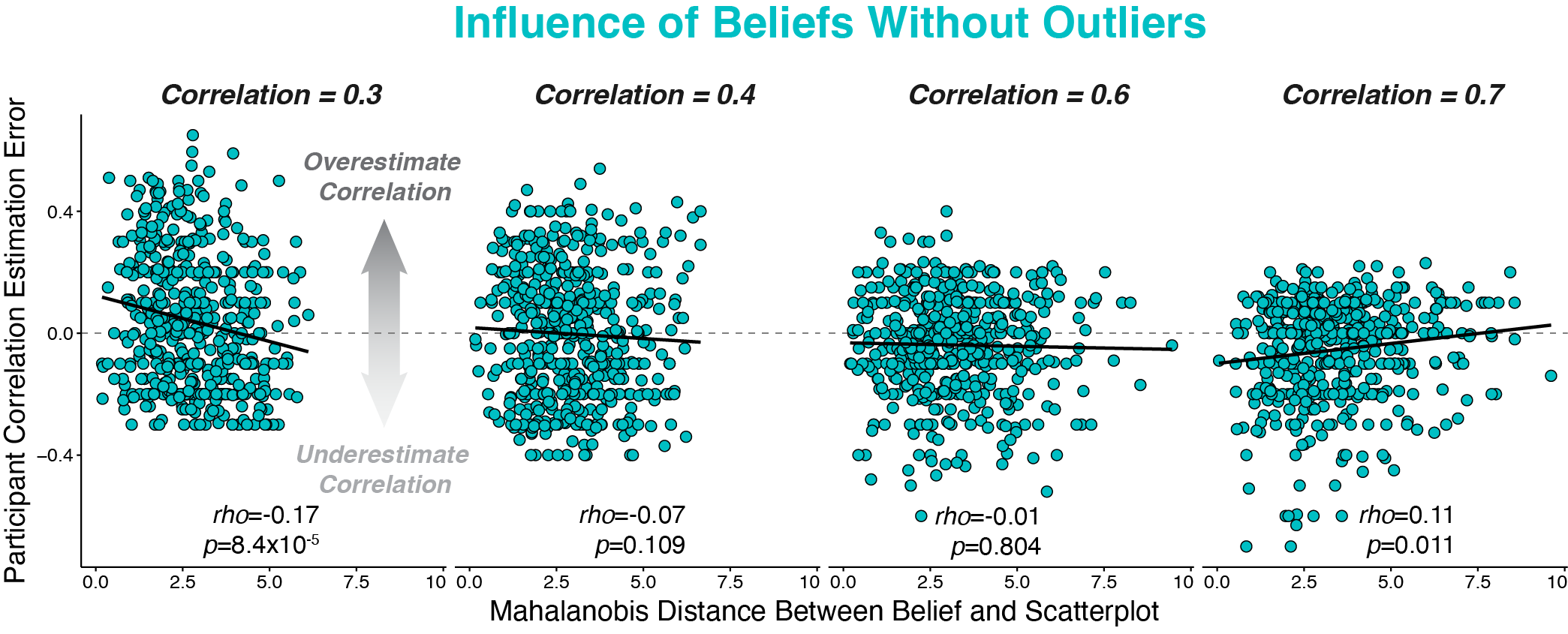}
        \caption{\label{fig5a}}
    \end{subfigure}
    \vspace{1pc}
    \begin{subfigure}[t]{0.7\textwidth}
        \includegraphics[width=\textwidth]{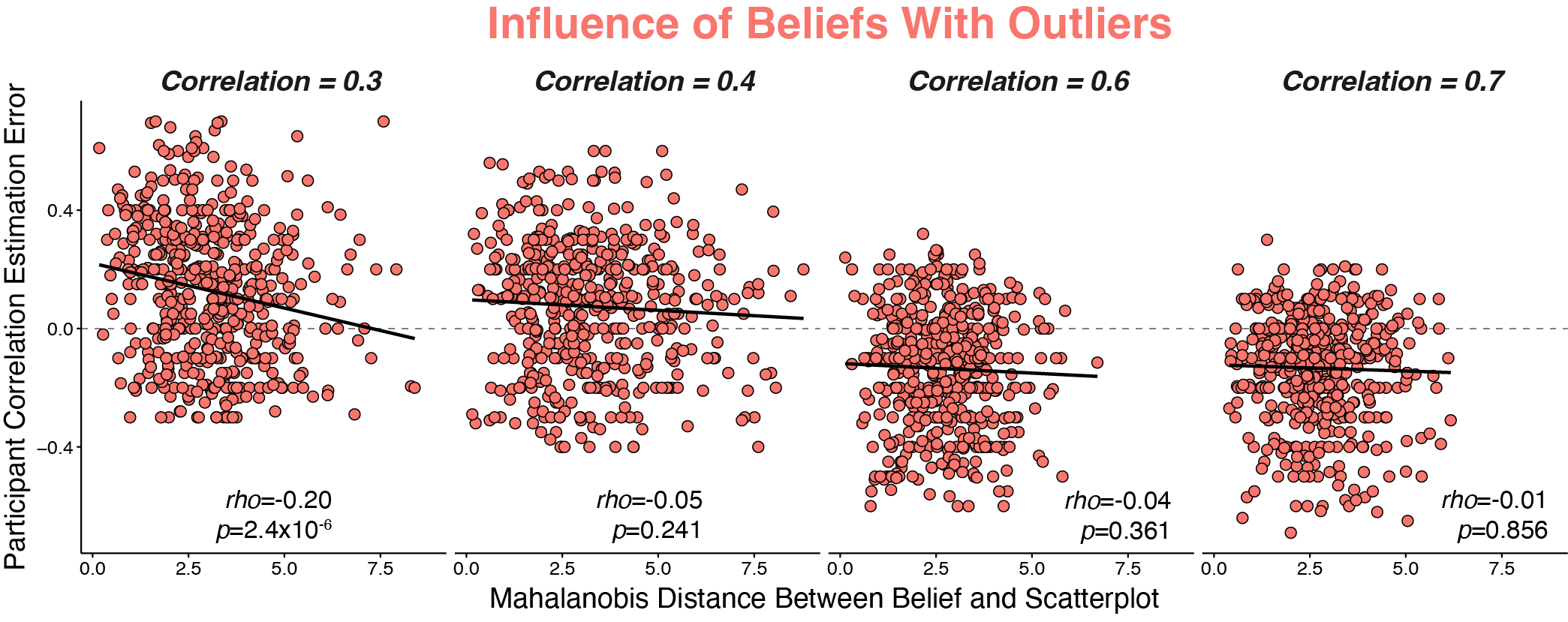}
        \caption{\label{fig5b}}
    \end{subfigure}
    \caption{\label{fig:fig5} Influence of beliefs on correlation estimations without outliers (cyan) and with outliers (pink). (a) On trials with weak generative correlations (correlation = 0.3) and no outliers, participants tended to overestimate the strength or trends that were close to their beliefs (i.e., confirmed their beliefs; low Mahalanobis distance) and underestimate the strength of trends that were far from their beliefs (i.e., went against their beliefs; high Mahalanobis distance). Distance from participant beliefs had no effect when generative correlations were 0.4 or above. (b) Same correlation plots as in (a) but on trials that include outliers. Beliefs had the same influence on correlation estimates, and for the same generative correlation values, as on trials without outliers. Trend lines correspond to ordinary least-squares regression lines. All reported statistics correspond to Spearman correlation values.}
\end{figure*}

Similar to Experiment 3, and irrespective of whether or not outliers were present in the scatter plots, beliefs had a slight influence on people's ratings of weak, but not strong correlations (Fig.~\ref{fig:fig5}). For weak correlations (i.e., correlations = 0.3), participants again tended to overestimate the strength of trends that confirmed their beliefs, and underestimate the strength of trends that went against their beliefs, both on labelled scatter plots that included outliers (Spearman correlation [95\% CI]: $\rho=-.20 [-.28,-.12], p<.001$; Fig.~\ref{fig:fig5}b) and did not include outliers ($\rho=-.17 [-.26,-.09], p<.001$; Fig.~\ref{fig:fig5}a). 

However, although beliefs affected correlation estimates whether outliers were present or absent, the presence of outliers did not interact with the effect of beliefs (bootstrap t-test comparing Spearman correlation difference between trials with and without outliers when correlations = 0.3: median bootstrapped correlation difference = -.03 [standard error = .06], $p=.700$). Thus, although beliefs did influence correlation estimates, the presence of outliers did not exacerbate or mitigate this influence.

We additionally observed a slight positive correlation between participant errors and distance from beliefs in the strongest correlation condition when outliers were not present (Spearman correlation [95\% CI]: $\rho=.11 [.03, .20], p=.011$; Fig.~\ref{fig:fig5}a). However, we believe this to be a spurious result, as this correlation was not found in the outlier condition ($\rho=-0.01, p=.856$; Fig.~\ref{fig:fig5}b) nor in Experiment 3 ($\rho=.03 [-.10, .08], p=.600$; Fig.~\ref{fig:fig4}c). Moreover, the observed $p$-value does not survive a correction for multiple comparisons (Bonferroni corrected $p$-value for Experiment 4 = 0.05/8 correlation tests = 0.006).

\section{Discussion}

Our results provide a systematic study of the influence of common visual elements and cognitive biases on interpretations of trends presented in scatter plots. Overall we highlight three primary findings.

First, we provide a conceptual replication of previous findings~\cite{correll2017regression,liu2021data} that (1) people are able to accurately report correlation trends in scatter plots, (2) accuracy depends on trend variance, (3) outliers bias people's interpretations of trends in the direction of the outliers, but also that (4) outliers are given less weight in people's estimates than other points in the plot. Importantly, our results extend previous findings to a new response modality, one in which participants use numerical estimates to provide their trend estimations, and do not require lines, annotations, or other potential confounds to provide a response.

Second, we find that trend lines bias people's perceptions of correlation strength, making them see correlations as stronger than they are in reality. This may be due to the salient nature of trend lines~\cite{wang2022makes}, which may catch people's attention to a larger degree than the underlying data points. This focus on trends may make people perceive slopes in correlation trends more easily and thus interpret trends with even slight slopes as being stronger than they actually are. Although trend lines could lead people to overestimate weak correlations, we also found that trend lines are an effective tool to mitigate the influence of outliers if the trend lines exclude outliers in their calculation. This could, again, be the result of people paying more attention to the saliency of the trend lines, which makes them further down-weigh outlying points.

Third, we find that people's pre-existing beliefs can influence their perceptions of trends in cases where the trends themselves are weak and uncertain. This finding replicates the results from Xiong and colleagues that beliefs bias correlation estimates~\cite{xiong2022seeing}, and also extends this previous work by demonstrating that the influence of belief biases occurs primarily when data conditions are uncertain. This finding fits well with Bayesian theories of perception, which show that people's priors (their beliefs) exert a stronger influence on people's perception of external events (evidence) when these events are ambiguous and difficult to interpret~\cite{gold2007neural, glaze2015normative, glaze2018bias, filipowicz2020pupil, murphy2021adaptive, krishnamurthy2017arousal}. Our results propose that these perceptual results can also affect interpretations of visualizations, exerting some influence when trends are weak, but less influence when trends are stronger. Indeed, previous work suggests that when people observe unambiguous trends that go against their beliefs (e.g., trends related to the severity of COVID-19 cases), people tend not to trust the source of the visualization rather than create a false percept of information presented in the graph itself~\cite{lee2021viral}. Nevertheless, our results propose that beliefs can influence interpretations, primarily under circumstances where trends are highly uncertain or ambiguous.

The implication of our findings is that there is a tension between information as presented in a graph versus a person's interpretation of a graph. This presents a dilemma for designers. If a designer is aware that one set of people reading a graph are, for example, likely to interpret a correlation as higher than its underlying data merit, does the designer then have an obligation to adjust the graph's layout? Similarly, if a system can sense bias in a particular user (e.g., via surveys, data logs, or worn sensors), does the system designer have an obligation to adjust how the system presents data so that its representation for a particular user more closely matches the objective data? On one hand, doing so may elicit a mental representation of a correlation in the biased observer that is closer to the actual correlation. On the other hand, the graph would look different for different users, which may cross ethical boundaries for some designers. This is a debatable issue, and a complete answer is well beyond the scope of any single research paper. Our stance is that what ultimately matters is the information as it is represented in the mind of each user.

Past work has shown that seemingly minor adjustments to the visual appearance of supposedly objective scalar rating interactors can have large impacts on people's ratings, and that these issues can be ameliorated with careful and subtle modifications~\cite{matejka2016effect}. It stands to reason that the interpretation of graphical information may require similarly subtle modifications to achieve an objective representation of data when seen through the lens of personal bias. Designers could also use animation to address these concerns~\cite{heer2007animated}, showing the personalized graph (adjusted for biases) first then transitioning to a non-personalized version. Interactive animations could be particularly useful to help users \emph{construct} or \emph{question} a mental model of the visualization~\cite{lee2015people}.

\begin{figure*}[h!]
  \centering
  \begin{tabular}{cccc}
    \includegraphics[width=0.22\textwidth]{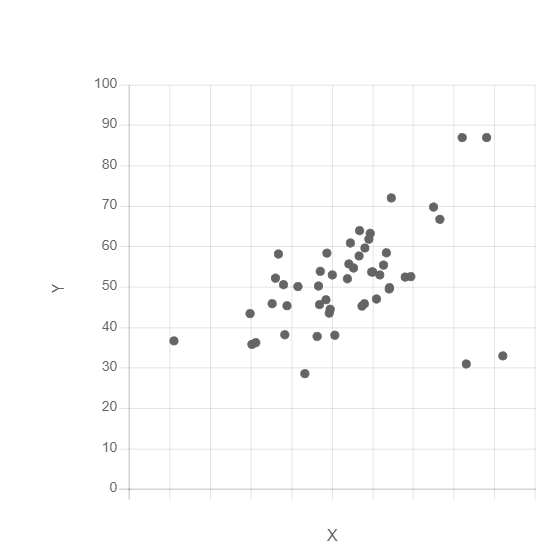} &
    \includegraphics[width=0.22\textwidth]{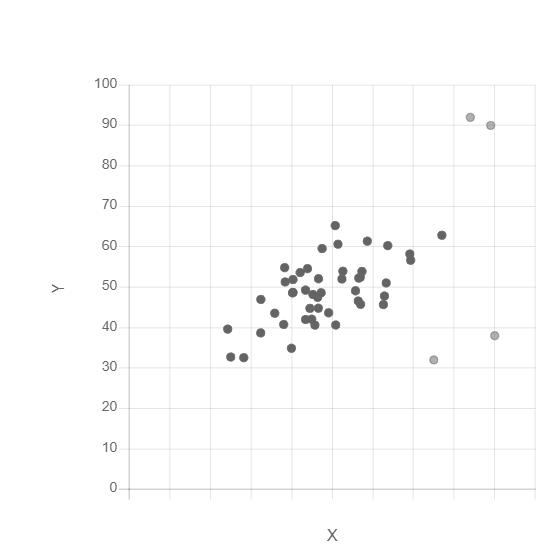} &
    \includegraphics[width=0.22\textwidth]{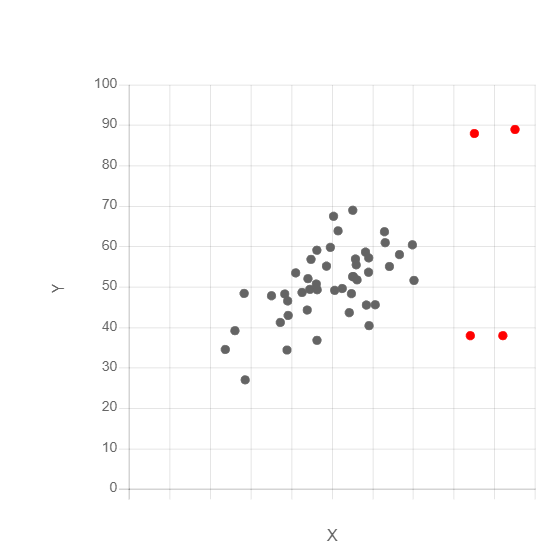} &
    \includegraphics[width=0.22\textwidth]{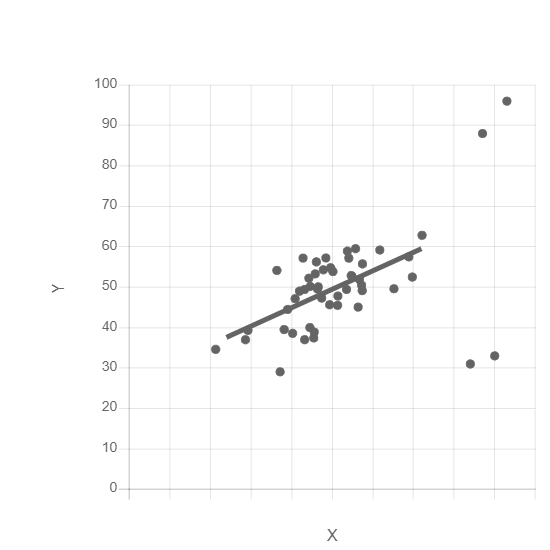}\\
    (a) & (b) & (c) & (d)\\
    
    \includegraphics[width=0.22\textwidth]{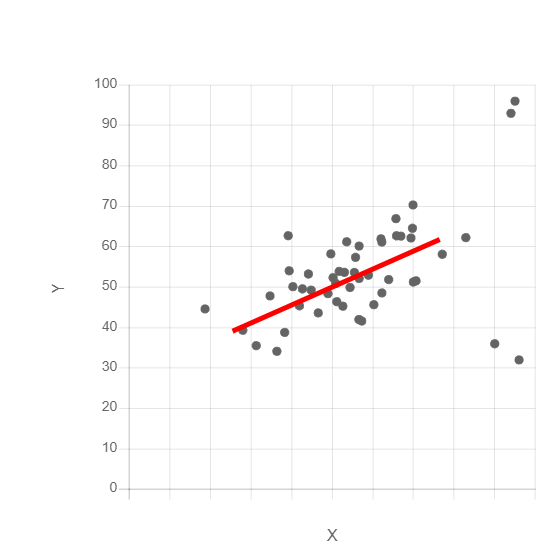} &
    \includegraphics[width=0.22\textwidth]{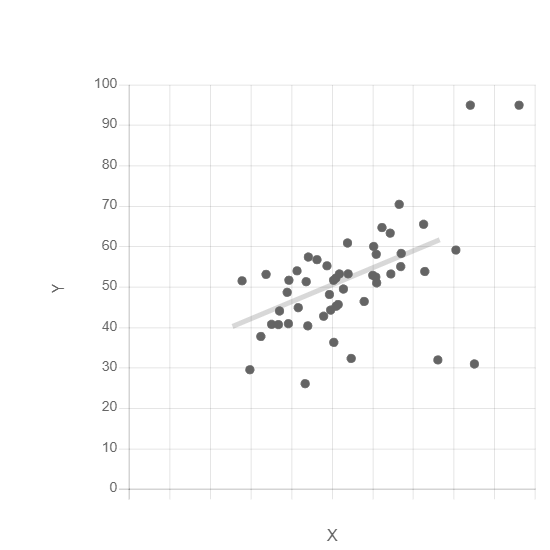} &
    \includegraphics[width=0.22\textwidth]{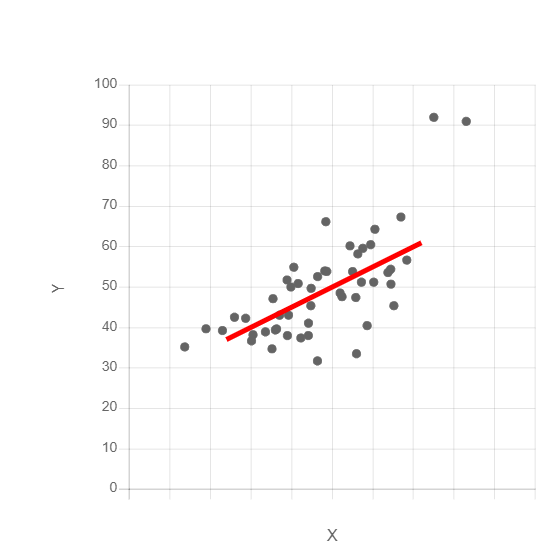} &
    \includegraphics[width=0.22\textwidth]{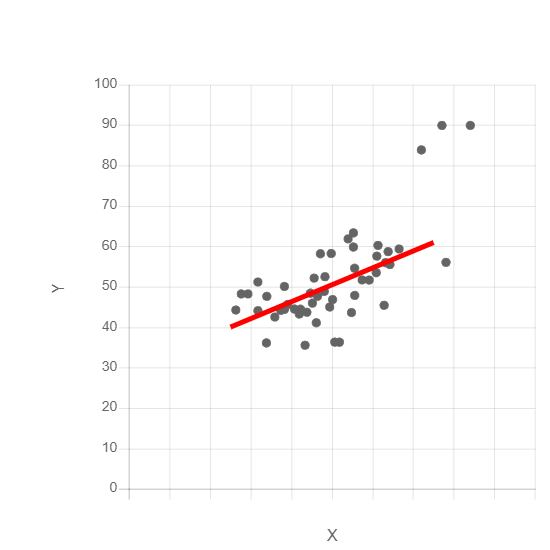} \\
    (e) & (f) & (g) & (h) \\ 
  \end{tabular}
  \caption{\label{fig:discussionGraphs} Different representations of a scatter plot showing the same amount of data with the same trend. The plot with no augmentation \textbf{(a)}; with outliers demphasized \textbf{(b)}; with outliers emphasized \textbf{(c)}; with a trend line \textbf{(d)}; with a trend line emphasized \textbf{(e)}; with a trend line de-emphasized \textbf{(f)}; with the trend line incorporating outliers \textbf{(g)}; with the trend line not incorporating outliers \textbf{(h)}.}
\end{figure*}

Still, this is a wicked problem, in part because biases and mental representations are difficult to sense with confidence. In the following sections, we discuss issues that our findings raise for designers. For each finding, we address the impact of adjusting the visual saliency of different components of a graph to direct user attention away from trends that a biased user might overemphasize or toward a trend that a biased user might underemphasize. Fig.~\ref{fig:discussionGraphs}a shows a scatter plot graph with no adjustments. Note in each case, these adjustments require some estimate of user bias, which could be derived from stated beliefs collected \emph{a priori} or implied beliefs from an analysis of e.g.\ a user's social media stream.

\begin{description}
\item \textbf{When a user has a strong bias in the opposite direction of a weak trend, they may tend to underestimate the trend itself}
When outliers are present, one ameliorating approach is to de-emphasize outliers, as in Fig.~\ref{fig:discussionGraphs}b. This has the visual effect of making the graph trend appear stronger. Note that this figure de-emphasizes outliers both in the \emph{direction of} and \emph{against} the trend. Another approach is to emphasize outliers when they are in the direction of the trend but not vice versa. However, there is a risk that users may interpret this as too manipulative, causing them to distrust the information presentation entirely. Adjusting the visual salience of \emph{all} outliers can ameliorate this concern. Another risk is that the system's understanding of user bias is incorrect or the system has low confidence in its estimate of user bias. To adjust, the system could adjust the impact of outlier emphasis/de-emphasis to match the system's confidence, provide manual controls to override incorrect biases, or use animation to toggle between the original and bias-adjusted graph.

Our work also shows that trend lines (such as the one shown in Figure~\ref{fig:discussionGraphs}d,e) will tend to make trends in scatter plots appear stronger than they actually are. For this reason, they can help align a biased user's interpretation of the graph when that user has strong beliefs in the opposite direction. 

\item \textbf{When users have a strong belief in the direction of a trend, they may overestimate the trend}  In this case, emphasizing outliers can bring attention to the relative lack of continuity of the overall trend (with respect to user bias), as in Fig.~\ref{fig:discussionGraphs}c. Opposite from the previous case, one could also emphasize outliers when they are against the trend but not vice versa. In addition to the risks and ameliorations already mentioned, designers might consider linking outliers to qualitative explanations (or exemplars) of data in the opposite direction of a trend.

Furthermore, since trend lines have such a considerable impact on people's perception of data, when users have strong biases in the direction of a trend, de-emphasizing the trend line (as in Fig.~\ref{fig:discussionGraphs}f) may more accurately convey the amplitude of that trend.

\item \textbf{Regardless of user beliefs, trend lines and outliers can change people's perception of data}
In particular, designers should keep in mind that a graph may more accurately convey a trend if the trend line \emph{excludes} outliers. Fig.~\ref{fig:discussionGraphs}g shows a graph with a trend line that incorporates outliers while in Fig.~\ref{fig:discussionGraphs}h the trend line is calculated without incorporating outliers. The difference appears subtle (and for that reason the trend line is shown in red in both graphs), but our findings reveal that even such minor adjustments to the presentation of a trend line can substantially impact how the user perceives the trend.
\end{description}


\subsection{Limitations and future work}

There are a number of limitations with our study that could be improved upon with future studies. First, we only studied elements that relate specifically to scatter plots, and it is unclear the extent to which our findings generalize to other visualization types. However, we do think our results can provide guidance for studies of different visualizations. Outliers can affect a number of different visualization types (e.g., line plots, box plots, violin plots) and visual elements meant to provide summaries, similar to trend lines, are also common (e.g., representations of means). Therefore, although our findings in their current form are limited to one visualization type, we see an opportunity for future research to understand the extent to which our findings generalize to other scenarios.

Second, it is not immediately clear how uncertainty needs to be represented in order for beliefs to exert an influence. In scatter plots, weak correlations are characterized by high variance in the position of the points. Uncertainty around the trend is represented by the variability of the elements a person is using to form their interpretation. These representations of uncertainty are closely tied to the circumstances under which human perceptual research find an influence of beliefs on perceptions of visual trends. However, other visualizations may represent uncertainty or trend weakness in a less ambiguous way (e.g., error bars in bar plots). Therefore, it is unclear the extent to which the influence of beliefs are related more to the ambiguity of the trend the person is trying to perceive, or to more general measures of uncertainty.

Last, although we did not observe interactions between beliefs and outliers, other scatter plot design elements (e.g., point shape, size, or transparency \cite{woodin2021conceptual}) could potentially play a role. These were out of scope for our study but could all be explored in future research.

Overall, we see these limitations as exciting opportunities for future research that could help improve our understanding of how users interpret different forms of visual information.

\section{Conclusion}
Despite their appearance of objectivity, graphs can have surprisingly subjective interpretations; people may look at the same graph or visualization and draw drastically different conclusions. Our investigation of one graph type, scatter plots, provided details about how visual features and cognitive biases impact people's interpretation of trends in these types of graphs. In sum, these findings imply that designers should carefully design graphical representations of statistical data, in particular making use of interactive and context-sensitive techniques to present data to help ameliorate misinterpretations and biases. We anticipate that our findings will contribute to a wide body of future work exploring other visual representations of data as well as design guidelines to mitigate the influence of biases.

\section{Data availability}
All supplemental material along with the data and code used in the creation of this article are available at \url{https://osf.io/76kp9/}.

\printbibliography 

@article{WANZER2021101896,
title = {The role of titles in enhancing data visualization},
journal = {Evaluation and Program Planning},
volume = {84},
pages = {101896},
year = {2021},
issn = {0149-7189},
doi = {https://doi.org/10.1016/j.evalprogplan.2020.101896},
url = {https://www.sciencedirect.com/science/article/pii/S0149718920302007},
author = {Dana Linnell Wanzer and Tarek Azzam and Natalie D. Jones and Darrel Skousen}
}

@article{2006-banking,
  title = {Multi-Scale Banking to 45 Degrees},
  author = {Jeffrey Heer AND Maneesh Agrawala},
  journal = {IEEE Trans. Visualization \& Comp. Graphics (Proc. InfoVis)},
  year = {2006},
  volume = {12},
  issue = {5},
  pages = {701--708},
  url = {http://vis.stanford.edu/papers/banking}
}

@ARTICLE{9195155,
  author={Henkin, Rafael and Turkay, Cagatay},
  journal={IEEE Transactions on Visualization and Computer Graphics}, 
  title={Words of Estimative Correlation: Studying Verbalizations of Scatterplots}, 
  year={2020},
  volume={},
  number={},
  pages={1-1},
  doi={10.1109/TVCG.2020.3023537}}

@ARTICLE{8305493,
author={Yang, Fumeng and Harrison, Lane T. and Rensink, Ronald A. and Franconeri, Steven L. and Chang, Remco},
journal={IEEE Transactions on Visualization and Computer Graphics},
title={Correlation Judgment and Visualization Features: A Comparative Study},
year={2019},
volume={25},
number={3},
pages={1474-1488},
doi={10.1109/TVCG.2018.2810918}}

@article{Bragdon2019UniversitySG,
  title={University students’ graph interpretation and comprehension abilities},
  author={Daniel Bragdon and Eric A. Pandiscio and Natasha M. Speer},
  journal={Investigations in Mathematics Learning},
  year={2019},
  volume={11},
  pages={275 - 290}
}

@article{heer2007animated,
  title={Animated transitions in statistical data graphics},
  author={Heer, Jeffrey and Robertson, George},
  journal={IEEE transactions on visualization and computer graphics},
  volume={13},
  number={6},
  pages={1240--1247},
  year={2007},
  publisher={IEEE}
}

@inproceedings{matejka2016effect,
  title={The effect of visual appearance on the performance of continuous sliders and visual analogue scales},
  author={Matejka, Justin and Glueck, Michael and Grossman, Tovi and Fitzmaurice, George},
  booktitle={Proceedings of the 2016 CHI Conference on Human Factors in Computing Systems},
  pages={5421--5432},
  year={2016}
}

@article{kaiser1962sample,
  title={Sample and population score matrices and sample correlation matrices from an arbitrary population correlation matrix},
  author={Kaiser, Henry F and Dickman, Kern},
  journal={Psychometrika},
  volume={27},
  number={2},
  pages={179--182},
  year={1962},
  publisher={Springer}
}

@article{Funkhouser1937,
  title={Historical development of the graphical representation of statistical data},
  author={Funkhouser, H Gray},
  journal={Osiris},
  volume={3},
  pages={269--404},
  year={1937},
  publisher={The Saint Catherine Press Ltd.}
}

@book{Playfair1801,
  title={The commercial and political atlas: representing, by means of stained copper-plate charts, the progress of the commerce, revenues, expenditure and debts of england during the whole of the eighteenth century},
  author={Playfair, William},
  year={1785},
  publisher={T. Burton}
}

@book{Cairo2019,
  title={How charts lie: Getting smarter about visual information},
  author={Cairo, Alberto},
  year={2019},
  publisher={WW Norton \& Company}
}

@article{palan2018prolific,
  title={Prolific. ac—A subject pool for online experiments},
  author={Palan, Stefan and Schitter, Christian},
  journal={Journal of Behavioral and Experimental Finance},
  volume={17},
  pages={22--27},
  year={2018},
  publisher={Elsevier}
}

@article{kuznetsova2017lmertest,
  title={lmerTest package: tests in linear mixed effects models},
  author={Kuznetsova, Alexandra and Brockhoff, Per B and Christensen, Rune HB},
  journal={Journal of statistical software},
  volume={82},
  number={1},
  pages={1--26},
  year={2017}
}

@article{lazer2018science,
  title={The science of fake news},
  author={Lazer, David MJ and Baum, Matthew A and Benkler, Yochai and Berinsky, Adam J and Greenhill, Kelly M and Menczer, Filippo and Metzger, Miriam J and Nyhan, Brendan and Pennycook, Gordon and Rothschild, David and others},
  journal={Science},
  volume={359},
  number={6380},
  pages={1094--1096},
  year={2018},
  publisher={American Association for the Advancement of Science}
}

@inproceedings{lee2021viral,
  title={Viral Visualizations: How Coronavirus Skeptics Use Orthodox Data Practices to Promote Unorthodox Science Online},
  author={Lee, Crystal and Yang, Tanya and Inchoco, Gabrielle D and Jones, Graham M and Satyanarayan, Arvind},
  booktitle={Proceedings of the 2021 CHI Conference on Human Factors in Computing Systems},
  pages={1--18},
  year={2021}
}

@incollection{cook2020introduction,
  title={Introduction to climate science denial},
  author={Cook, John},
  booktitle={Research Handbook on Communicating Climate Change},
  year={2020},
  publisher={Edward Elgar Publishing}
}

@article{nickerson1998confirmation,
  title={Confirmation bias: A ubiquitous phenomenon in many guises},
  author={Nickerson, Raymond S},
  journal={Review of general psychology},
  volume={2},
  number={2},
  pages={175--220},
  year={1998},
  publisher={SAGE Publications Sage CA: Los Angeles, CA}
}

@article{bates2014fitting,
  title={Fitting linear mixed-effects models using lme4},
  author={Bates, Douglas and M{\"a}chler, Martin and Bolker, Ben and Walker, Steve},
  journal={arXiv preprint arXiv:1406.5823},
  year={2014}
}

@article{jabar2017tuned,
  title={Tuned by experience: How orientation probability modulates early perceptual processing},
  author={Jabar, Syaheed B and Filipowicz, Alexandre and Anderson, Britt},
  journal={Vision Research},
  volume={138},
  pages={86--96},
  year={2017},
  publisher={Elsevier}
}

@article{glaze2015normative,
  title={Normative evidence accumulation in unpredictable environments},
  author={Glaze, Christopher M and Kable, Joseph W and Gold, Joshua I},
  journal={Elife},
  volume={4},
  pages={e08825},
  year={2015},
  publisher={eLife Sciences Publications Limited}
}

@article{gold2007neural,
  title={The neural basis of decision making},
  author={Gold, Joshua I and Shadlen, Michael N},
  journal={Annu. Rev. Neurosci.},
  volume={30},
  pages={535--574},
  year={2007},
  publisher={Annual Reviews}
}

@article{stottinger2006dissociating,
  title={Dissociating size representation for action and for conscious judgment: Grasping visual illusions without apparent obstacles},
  author={St{\"o}ttinger, Elisabeth and Perner, Josef},
  journal={Consciousness and Cognition},
  volume={15},
  number={2},
  pages={269--284},
  year={2006},
  publisher={Elsevier}
}

@inproceedings{battle2018beagle,
  title={Beagle: Automated extraction and interpretation of visualizations from the web},
  author={Battle, Leilani and Duan, Peitong and Miranda, Zachery and Mukusheva, Dana and Chang, Remco and Stonebraker, Michael},
  booktitle={Proceedings of the 2018 CHI Conference on Human Factors in Computing Systems},
  pages={1--8},
  year={2018}
}

@article{de2011robust,
  title={Robust averaging during perceptual judgment},
  author={De Gardelle, Vincent and Summerfield, Christopher},
  journal={Proceedings of the National Academy of Sciences},
  volume={108},
  number={32},
  pages={13341--13346},
  year={2011},
  publisher={National Acad Sciences}
}

@article{wei2015bayesian,
  title={A Bayesian observer model constrained by efficient coding can explain'anti-Bayesian'percepts},
  author={Wei, Xue-Xin and Stocker, Alan A},
  journal={Nature neuroscience},
  volume={18},
  number={10},
  pages={1509--1517},
  year={2015},
  publisher={Nature Publishing Group}
}

@article{filipowicz2018rejecting,
  title={Rejecting outliers: Surprising changes do not always improve belief updating.},
  author={Filipowicz, Alexandre and Valadao, Derick and Anderson, Britt and Danckert, James},
  journal={Decision},
  volume={5},
  number={3},
  pages={165},
  year={2018},
  publisher={Educational Publishing Foundation}
}

@article{summerfield2015humans,
  title={Do humans make good decisions?},
  author={Summerfield, Christopher and Tsetsos, Konstantinos},
  journal={Trends in cognitive sciences},
  volume={19},
  number={1},
  pages={27--34},
  year={2015},
  publisher={Elsevier}
}

@article{kalish2007iterated,
  title={Iterated learning: Intergenerational knowledge transmission reveals inductive biases},
  author={Kalish, Michael L and Griffiths, Thomas L and Lewandowsky, Stephan},
  journal={Psychonomic Bulletin \& Review},
  volume={14},
  number={2},
  pages={288--294},
  year={2007},
  publisher={Springer}
}

@Manual{rcore,
  title = {R: A Language and Environment for Statistical Computing},
  author = {{R Core Team}},
  organization = {R Foundation for Statistical Computing},
  address = {Vienna, Austria},
  year = {2021},
  url = {https://www.R-project.org/},
  }

@article{glaze2018bias,
  title={A bias--variance trade-off governs individual differences in on-line learning in an unpredictable environment},
  author={Glaze, Christopher M and Filipowicz, Alexandre LS and Kable, Joseph W and Balasubramanian, Vijay and Gold, Joshua I},
  journal={Nature Human Behaviour},
  volume={2},
  number={3},
  pages={213--224},
  year={2018},
  publisher={Nature Publishing Group}
}

@article{filipowicz2020pupil,
  title={Pupil diameter encodes the idiosyncratic, cognitive complexity of belief updating},
  author={Filipowicz, Alexandre LS and Glaze, Christopher M and Kable, Joseph W and Gold, Joshua I},
  journal={Elife},
  volume={9},
  pages={e57872},
  year={2020},
  publisher={eLife Sciences Publications Limited}
}

@article{murphy2021adaptive,
  title={Adaptive circuit dynamics across human cortex during evidence accumulation in changing environments},
  author={Murphy, Peter R and Wilming, Niklas and Hernandez-Bocanegra, Diana C and Prat-Ortega, Genis and Donner, Tobias H},
  journal={Nature Neuroscience},
  pages={1--11},
  year={2021},
  publisher={Nature Publishing Group}
}

@article{krishnamurthy2017arousal,
  title={Arousal-related adjustments of perceptual biases optimize perception in dynamic environments},
  author={Krishnamurthy, Kamesh and Nassar, Matthew R and Sarode, Shilpa and Gold, Joshua I},
  journal={Nature human behaviour},
  volume={1},
  number={6},
  pages={1--11},
  year={2017},
  publisher={Nature Publishing Group}
}

@article{mccright2013influence,
  title={The influence of political ideology on trust in science},
  author={McCright, Aaron M and Dentzman, Katherine and Charters, Meghan and Dietz, Thomas},
  journal={Environmental Research Letters},
  volume={8},
  number={4},
  pages={044029},
  year={2013},
  publisher={IOP Publishing}
}

@article{baron2019actively,
  title={Actively open-minded thinking in politics},
  author={Baron, Jonathan},
  journal={Cognition},
  volume={188},
  pages={8--18},
  year={2019},
  publisher={Elsevier}
}

@article{caprara2005new,
  title={A new scale for measuring adults' prosocialness},
  author={Caprara, Gian Vittorio and Steca, Patrizia and Zelli, Arnaldo and Capanna, Cristina},
  journal={European Journal of psychological assessment},
  volume={21},
  number={2},
  pages={77--89},
  year={2005},
  publisher={Hogrefe \& Huber Publishers}
}

@article{mildenberger2019beliefs,
  title={Beliefs about climate beliefs: the importance of second-order opinions for climate politics},
  author={Mildenberger, Matto and Tingley, Dustin},
  journal={British Journal of Political Science},
  volume={49},
  number={4},
  pages={1279--1307},
  year={2019},
  publisher={Cambridge University Press}
}

@article{lee2015people,
  title={How do people make sense of unfamiliar visualizations?: A grounded model of novice's information visualization sensemaking},
  author={Lee, Sukwon and Kim, Sung-Hee and Hung, Ya-Hsin and Lam, Heidi and Kang, Youn-ah and Yi, Ji Soo},
  journal={IEEE transactions on visualization and computer graphics},
  volume={22},
  number={1},
  pages={499--508},
  year={2015},
  publisher={IEEE}
}

@article{harrison2014corrweber,
  title={Ranking visualizations of correlation using weber's law},
  author={Harrison, Lane and Yang, Fumeng and Franconeri, Steven and Chang, Remco},
  journal={IEEE transactions on visualization and computer graphics},
  volume={20},
  number={12},
  pages={1943--1952},
  year={2014},
  publisher={IEEE}
}

@inproceedings{correll2017regression,
  title={Regression by eye: Estimating trends in bivariate visualizations},
  author={Correll, Michael and Heer, Jeffrey},
  booktitle={Proceedings of the 2017 CHI Conference on Human Factors in Computing Systems},
  pages={1387--1396},
  year={2017}
}

@inproceedings{kim2017explaining,
    title={Explaining the gap: Visualizing one's predictions improves recall and comprehension of data},
    author={Kim, Yea-Seul and Reinecke, Katharina and Hullman, Jessica},
    booktitle={Proceedings of the 2017 CHI Conference on Human Factors in Computing Systems},
    pages={1375--1386},
    year={2017}
}

@inproceedings{liu2021data,
  title={Data-Driven Mark Orientation for Trend Estimation in Scatterplots},
  author={Liu, Tingting and Li, Xiaotong and Bao, Chen and Correll, Michael and Tu, Changehe and Deussen, Oliver and Wang, Yunhai},
  booktitle={Proceedings of the 2021 CHI Conference on Human Factors in Computing Systems},
  pages={1--16},
  year={2021}
}

@article{pandey2014persuasive,
  title={The persuasive power of data visualization},
  author={Pandey, Anshul Vikram and Manivannan, Anjali and Nov, Oded and Satterthwaite, Margaret and Bertini, Enrico},
  journal={IEEE transactions on visualization and computer graphics},
  volume={20},
  number={12},
  pages={2211--2220},
  year={2014},
  publisher={IEEE}
}

@article{woodin2021conceptual,
  title={Conceptual metaphor and graphical convention influence the interpretation of line graphs},
  author={Woodin, Greg and Winter, Bodo and Padilla, Lace},
  journal={IEEE transactions on visualization and computer graphics},
  volume={28},
  number={2},
  pages={1209--1221},
  year={2021},
  publisher={IEEE}
}

@inproceedings{rensink2010perception,
  title={The perception of correlation in scatterplots},
  author={Rensink, Ronald A and Baldridge, Gideon},
  booktitle={Computer Graphics Forum},
  volume={29},
  number={3},
  pages={1203--1210},
  year={2010},
  organization={Wiley Online Library}
}

@inproceedings{li2009evaluation,
  title={Evaluation of symbol contrast in scatterplots},
  author={Li, Jing and van Wijk, Jarke J and Martens, Jean-Bernard},
  booktitle={2009 IEEE Pacific visualization symposium},
  pages={97--104},
  year={2009},
  organization={IEEE}
}

@inproceedings{smart2019measuring,
  title={Measuring the separability of shape, size, and color in scatterplots},
  author={Smart, Stephen and Szafir, Danielle Albers},
  booktitle={Proceedings of the 2019 CHI Conference on Human Factors in Computing Systems},
  pages={1--14},
  year={2019}
}

@article{tremmel1995visual,
  title={The visual separability of plotting symbols in scatterplots},
  author={Tremmel, Lothar},
  journal={Journal of Computational and Graphical Statistics},
  volume={4},
  number={2},
  pages={101--112},
  year={1995},
  publisher={Taylor \& Francis}
}

@article{wang2022makes,
  title={What makes a scatterplot hard to comprehend: data size and pattern salience matter},
  author={Wang, Jiachen and Cai, Xiwen and Su, Jiajie and Liao, Yu and Wu, Yingcai},
  journal={Journal of Visualization},
  volume={25},
  number={1},
  pages={59--75},
  year={2022},
  publisher={Springer}
}

@article{szafir2016four,
  title={Four types of ensemble coding in data visualizations},
  author={Szafir, Danielle Albers and Haroz, Steve and Gleicher, Michael and Franconeri, Steven},
  journal={Journal of vision},
  volume={16},
  number={5},
  pages={11--11},
  year={2016},
  publisher={The Association for Research in Vision and Ophthalmology}
}

@article{ariely2001seeing,
  title={Seeing sets: Representation by statistical properties},
  author={Ariely, Dan},
  journal={Psychological science},
  volume={12},
  number={2},
  pages={157--162},
  year={2001},
  publisher={SAGE Publications Sage CA: Los Angeles, CA}
}

@article{ciccione2022outlier,
  title={\change{Outlier detection and rejection in scatterplots: Do outliers influence intuitive statistical judgments?}},
  author={Ciccione, Lorenzo and Dehaene, Guillaume and Dehaene, Stanislas},
  year={2022},
  publisher={PsyArXiv}
}

@article{dimara2018task,
  title={\change{A task-based taxonomy of cognitive biases for information visualization}},
  author={Dimara, Evanthia and Franconeri, Steven and Plaisant, Catherine and Bezerianos, Anastasia and Dragicevic, Pierre},
  journal={IEEE transactions on visualization and computer graphics},
  volume={26},
  number={2},
  pages={1413--1432},
  year={2018},
  publisher={IEEE}
}

@article{micallef2017towards,
  title={\change{Towards perceptual optimization of the visual design of scatterplots}},
  author={Micallef, Luana and Palmas, Gregorio and Oulasvirta, Antti and Weinkauf, Tino},
  journal={IEEE transactions on visualization and computer graphics},
  volume={23},
  number={6},
  pages={1588--1599},
  year={2017},
  publisher={IEEE}
}

@inproceedings{wall2017warning,
  title={\change{Warning, bias may occur: A proposed approach to detecting cognitive bias in interactive visual analytics}},
  author={Wall, Emily and Blaha, Leslie M and Franklin, Lyndsey and Endert, Alex},
  booktitle={IEEE conference on visual analytics science and technology (VAST)},
  pages={104--115},
  year={2017},
  organization={IEEE}
}

@article{reimann2021visual,
  title={\change{Visual Model Fit Estimation in Scatterplots: Influence of Amount and Decentering of Noise}},
  author={Reimann, Daniel and Blech, Christine and Ram, Nilam and Gaschler, Robert},
  journal={IEEE Transactions on Visualization and Computer Graphics},
  volume={27},
  number={9},
  pages={3834--3838},
  year={2021},
  publisher={IEEE}
}

@article{xiong2022seeing,
  title={\change{Seeing what you believe or believing what you see? belief biases correlation estimation}},
  author={Xiong, Cindy and Stokes, Chase and Kim, Yea-Seul and Franconeri, Steven},
  journal={IEEE Transactions on Visualization and Computer Graphics},
  year={2022},
  publisher={IEEE}
}

\end{document}